\documentclass[sn-basic]{sn-jnl}

\usepackage{graphicx}%
\usepackage{multirow}%
\usepackage{amsmath,amssymb,amsfonts}%
\usepackage{amsthm}%
\usepackage{mathrsfs}%
\usepackage[title]{appendix}%
\usepackage{xcolor}%
\usepackage{textcomp}%
\usepackage{manyfoot}%
\usepackage{booktabs}%
\usepackage{algorithm}%
\usepackage{algorithmicx}%
\usepackage{algpseudocode}%
\usepackage{listings}%
\usepackage[utf8x]{inputenc}
\usepackage{siunitx}

\newcommand{\red}[1]{\textcolor{black}{#1}}

\def \bp{$\beta$\,Pic}

\begin{document}

\title[Article Title]{The evolution of exocomets and their source populations}


\author*[1]{\fnm{Alexander J.} \sur{Mustill}}\email{alexander.mustill@fysik.lu.se}

\author[2]{\fnm{Tim} \sur{Pearce}}\email{tim.pearce@warwick.ac.uk}
\equalcont{These authors contributed equally to this work.}

\author[3]{\fnm{Michele} \sur{Bannister}}\email{michele.bannister@canterbury.ac.nz}
\equalcont{These authors contributed equally to this work.}

\author[4]{\fnm{Susanne} \sur{Pfalzner}}\email{s.pfalzner@fz-juelich.de}
\equalcont{These authors contributed equally to this work.}

\author[5,19]{\fnm{Dag} \sur{Evensberget}}\email{dag.evensberget@geo.uio.no}

\author[2,6,7]{\fnm{Dimitri} \sur{Veras}}\email{dimitri.veras@aya.yale.edu}

\author[8,9]{\fnm{Rosita} \sur{Kokotanekova}}\email{rkokotanekova@nao-rozhen.org}

\author[3,10]{\fnm{Matthew} \sur{Hopkins}}\email{matthewj.hopkins@canterbury.ac.nz}

\author[11]{\fnm{Dennis} \sur{Bodewits}}\email{dennis@auburn.edu}

\author[12,13]{\fnm{Darryl Z.}
\sur{Seligman}}\email{dzs@msu.edu}

\author[14]{\fnm{Isabel} \sur{Rebollido}}\email{isabel.rebollidovazquez@esa.int}

\author[15,16]{\fnm{Raphael}\sur{Marschall}}\email{raphael.marschall@unibe.ch}

\author[17]{\fnm{Bin} \sur{Yang}}\email{bin.yang@mail.udp.cl}

\author[1]{\fnm{Klaudia} \sur{Jaworska}}\email{kl8058ja-s@student.lu.se}

\author[18]{\fnm{Xabier} \sur{P\'erez Couto}}\email{xabier.perez.couto@udc.gal}

\affil*[1]{\orgdiv{Department of Physics}, \orgname{Lund University}, \orgaddress{\street{Box 118}, \city{Lund}, \postcode{22100}, \country{Sweden}}}

\affil[2]{\orgdiv{Department of Physics}, \orgname{University of Warwick}, \orgaddress{\street{Gibbet Hill Road}, \city{Coventry}, \postcode{CV4 7AL}, \country{United Kingdom}}}

\affil[3]{\orgdiv{School of Physical and Chemical Sciences -- Te Kura Mat\={u}, \orgname{University of Canterbury}, \orgaddress{\street{Private Bag 4800}, \city{Christchurch}, \postcode{8140}, \country{Aotearoa New Zealand}}}}

\affil[4]{\orgdiv{J\"ulich Supercomputing Centre}, \orgname{Research Centre J\"ulich}, \orgaddress{\street{Wilhelm-Johnen-Strasse}, \city{J\"ulich}, \postcode{52428}, \country{Germany}}}

\affil[5]{\orgdiv{Leiden Observatory}, \orgname{Leiden University}, \orgaddress{\street{PO Box 9513}, \city{Leiden}, \postcode{2300~RA}, \country{The Netherlands}}}

\affil[6]{\orgdiv{Centre for Exoplanets and Habitability}, \orgname{University of Warwick}, \orgaddress{\street{Gibbet Hill Road}, \city{Coventry}, \postcode{CV4 7AL}, \country{United Kingdom}}}

\affil[7]{\orgdiv{Centre for Space Domain Awareness}, \orgname{University of Warwick}, \orgaddress{\street{Gibbet Hill Road}, \city{Coventry}, \postcode{CV4 7AL}, \country{United Kingdom}}}


\affil[8]{\orgdiv{Institute of Astronomy and NAO}, \orgname{Bulgarian Academy of Sciences}, \orgaddress{\street{72 Tsarigradsko Shose Blvd.}, \city{Sofia}, \postcode{1784}, \country{Bulgaria}}}

\affil[9]{\orgname{International Space Science Institute}, \orgaddress{\street{Hallerstrasse 6}, \city{Bern}, \postcode{3012}, \country{Switzerland}}}

\affil[10]{\orgdiv{Department of Physics}, \orgname{University of Oxford}, \orgaddress{\street{Keble Road}, \city{Oxford}, \postcode{OX1 3RH}, \country{United Kingdom}}}


\affil[11]{\orgdiv{Department of Physics}, \orgname{Auburn University}, \orgaddress{\street{Edmund C. Leach Science Center}, \city{Auburn}, \postcode{36849}, \state{AL} \country{USA}}}

\affil[12]{\orgdiv{Department of Physics and Astronomy}, \orgname{Michigan State University}, \orgaddress{\city{East Lansing}, \postcode{MI 48824},  \country{USA}}}

\affil[13]{\orgdiv{NSF Astronomy and Astrophysics Postdoctoral Fellow}}

\affil[14]{\orgdiv{European Space Astronomy Centre (ESAC)}, \orgname{European Space Agency}, \orgaddress{\street{Camino Bajo del Castillo s/n}, \city{Villanueva de la Cañada}, \postcode{28692}, \state{Madrid} \country{Spain}}}


\affil[15]{\orgdiv{Laboratoire J.-L. Lagrange}, \orgname{Observatoire de la C\^ote d'Azur, CNRS}, \orgaddress{\street{CS 34229}, \city{Nice Cedex 4}, \postcode{06304}, \country{France}}}

\affil[16]{\orgdiv{Physics Institute, Space Research and Planetary Sciences}, \orgname{University of Bern}, \orgaddress{\street{Sidlerstrasse 5}, \city{Bern}, \postcode{CH-3012}, \country{Switzerland}}}

\affil[17]{\orgdiv{Instituto de Estudios Astrofísicos, Facultad de Ingeniería y Ciencias}, \orgname{Universidad Diego Portales}, \orgaddress{\street{Av. Ej\'ercito Libertador 441}, \city{Santiago},  \postcode{8370191}, \country{Chile}}}

\affil[18]{\orgdiv{Centre for Research in Information and Communication Technologies}, \orgname{Universidade da Coruña}, \orgaddress{\street{Campus de Elviña s/n}, \city{A Coruña}, \postcode{15071}, \state{Galiza} \country{Spain}}}

\affil[19]{\orgdiv{Centre for Planetary Habitability (PHAB), Department of Geosciences}, \orgname{University of Oslo}, 
\orgaddress{\street{Sem Sælands vei 2A}, \city{Oslo}, \postcode{0371}, \country{Norway}}}

\abstract{We review the current state of knowledge of the long-term evolution of the small bodies that give rise to comets and exocomets, as well as their reservoirs. The active cometary phase is only transitory, and bodies that become comets pass from a source population, such as the Kuiper Belt, Oort cloud or their extra-solar analogues, through the active cometary phase, to eventual dormancy or destruction. We discuss dynamical delivery channels that can move comets from their source reservoirs to orbits with small periapsides, and the depletion of these reservoirs by dynamical and collisional means. We also discuss the physical evolution of cometary nuclei, especially in light of recent advances from missions to Solar System comets such as Rosetta's visit to 67P. We then describe our current knowledge of interstellar objects, which can originate from the same source regions as exocomets but be amenable to detailed study when they enter the Solar System. We include a summary of stellar winds emanating from different types of stars, which become increasingly strong once stars leave the Main Sequence. This is followed by a description of how small bodies are affected by stellar evolution, and the range of comet-like phenomena observed towards white dwarf stars. Overall, while we have an increasingly good picture of the physical and dynamical evolution of Solar System comets, a number of large gaps remain in our knowledge of the physics of exocomets, related to our inability to directly probe these bodies and many of the planets that might be affecting their orbits.}

\keywords{Comets --- Kuiper Belt Objects --- Oort Cloud --- planets and satellites: dynamical evolution and stability --- planet–disc interactions --- circumstellar matter} 



\maketitle

\section{Introduction}
\label{sec:intro}

An exocomet is a small body orbiting another star that is undergoing detectable mass loss, either through sublimation or by some other means\footnote{This could include tidal disruption, as is likely the case for some bodies orbiting white dwarfs.} \citep{Iglesias2025}. Comets and other small bodies of planetary systems are constantly evolving, both as isolated bodies and as populations. Most clearly, when a small body presents as an active comet it is, by definition, losing mass which becomes visible as its coma and tails; this implies that a body's existence as an active comet cannot continue indefinitely. The mass loss arises because of changing conditions as the cometary nucleus approaches orbital periapsis, notably the increased instellation 
which heats the body and leads to outgassing. As an orbit in the Newtonian two-body problem is invariant, this in turn implies that a comet's orbit must have been changed in the relatively recent past so that it has not spent too long with its current, small, orbital periapsis. This in turn implies the existence of one or more cometary reservoirs where the comets are ``stored'' for long periods (up to Gyrs) at large radii from their host star, and some perturbing forces that can remove them from these reservoirs to be driven onto orbits with small periapsides where they become visible as active comets.

In the Solar System, these delivery processes are reasonably well understood, with distinct dynamical populations of comets being known, together with the means by which they are removed from their source reservoirs of the Kuiper Belt, Scattered Disc and Oort Cloud \citep{Dones2015,Peixinho2020}. The two major populations of comets are (1) ecliptic comets, which are moved by perturbations induced by the giant planets from their source in the trans-Neptunian disc (Kuiper Belt proper and scattered disc), via the Centaur population, to become Jupiter-Family Comets, and (2) the nearly-isotropic comets, which have been scattered into the Oort cloud by the giant planets earlier in the Solar System's history, and later re-appear with small perihelia following perturbations to their orbits induced by nearby stars and the Galactic tide. Further details and subtleties of this dynamical classification can be found in, e.g., \cite{Dones2015}. Over the past few decades we have also made great strides in characterising both the populations and the dynamical and collisional conditions within the Solar System's source reservoirs, with the number of known Trans-Neptunian Objects (TNOs) growing from one (the binary Pluto--Charon system) in 1990 to many thousands at present. The detected TNOs, however, are much larger than typical cometary nuclei owing to detection biases: we still lack a census of TNOs at sizes below $\sim100$\,km, while the Oort Cloud remains largely hypothetical and undetected, save for those few members that have intruded into the inner Solar System\footnote{Specifically, we mean interior to the orbit of Jupiter, although some comets do show activity at larger heliocentric distances.}, owing to the immense distances ($\gtrsim1000$\,au) at which bodies are thought to reside there.

The situation is much less understood when it comes to small body and cometary phenomena outside the Solar System. We now have spectroscopic and photometric evidence for exocomets transiting a handful of other Main Sequence stars, notably $\beta$ Pic (\citealt{Ferlet1987,LagrangeHenri1988}, and reviewed extensively by Lu et al., this issue). A large fraction of Main Sequence stars exhibit exo-zodiacal light or ``exozodi'', a near- or mid-IR excess located at $\lesssim1$\,au whose origin is uncertain but may in part arise from dust released by comets near perihelion \citep{Beichman2005,Absil2013,Ertel2020}. Potential source populations are reasonably well-characterised, at least when it comes to more massive analogues of the Kuiper Belt which are easily visible as mid- or far-IR excess, or resolved thermal emission or scattered light, in the form of debris discs \citep{Aumann1984,Smith1984}. On the other hand, the dynamics of delivery from debris disc body to comet is much less clear than it is in the Solar System. This is because the bodies responsible for delivery are nearly all undetectable with current technology, since most exoplanet detection methods favour closer orbits $\lesssim$ few au, and all favour only more massive planets. While we have a good understanding of stellar multiplicity \citep{Raghavan2010}, and some census of super-Jupiter planets on wide orbits from direct imaging surveys \citep{Bowler2016,Nielsen2019,Vigan2021}, the demographics of smaller planets at tens of au are mostly unconstrained. We have learned from the explosion of exoplanet discoveries in the past three decades that planetary systems are common, diverse, and often do not at all resemble the Solar System \citep{Winn2015,Zhu2021}. For exocomets this situation is both vexing, since we know little for certain about what might be responsible for delivery, and inspiring, since a great range of planetary architectures and dynamical mechanisms can be considered. Only in the case of $\beta$ Pic do we have a good enough characterisation of the planets, the cometary activity, and the source reservoir; this unique system is discussed in detail in Lu et al., this issue.

In this work, we discuss the long-term changes experienced by cometary reservoirs, and the small bodies that become comets themselves, inside and outside the Solar System. We discuss processes that take place after the epoch of formation in the protoplanetary disc, which is discussed in \cite{Bannister2025}. Our discussion of the physical evolution of cometary nuclei focuses on long-term processes; for a detailed discussion of the physics of mass loss from the nucleus and the formation of the comae and tails on (sub-)orbital timescales, see \citet[this issue]{Vrignaud2026}. We begin the work by discussing the different dynamical mechanisms that have been proposed to move bodies out of their debris disc reservoirs to small periapsides (Section~\ref{sec: mechanismsToGetCometsCloseToStars}). We then discuss the long-term evolution of the source reservoirs, both by the aforementioned dynamical processes as well as by collisions between small bodies, which are unimportant in the present TNO and Oort Cloud populations in the Solar System but are significant in observable extra-solar debris discs (Section~\ref{sec:depletion}). Following that, we discuss the dynamics of the Oort Cloud and exo-Oort Clouds (Section~\ref{sec:Oort}), where previously we focused on delivery from flatter disc populations.
After that, we briefly discuss our state of knowledge surrounding direct observations that can probe these dynamical links in specific cases (Section~\ref{sec:observations}).

We then switch our attention to discuss the evolution of small bodies as individual objects. In Section~\ref{sec: nonGravAndTidalForces} we discuss nongravitational forces acting on small bodies to change their orbits and spin states, as well as tidal disruption as an ultimate end state.  
We then discuss the physical processes affecting small bodies in themselves, rather than their orbits, discussing storage and the transition to activity, followed by their death or dormancy (Section~\ref{sec:evolution}). These processes are best studied for the Solar System, where a succession of space missions have visited comets to study them as geological bodies.
We follow this with a description of the three hitherto-detected interstellar objects (ISOs), both as individual objects, and looking ahead what we might learn from statistical studies of the Galactic kinematics of ISOs about minor body formation in different stellar populations (Section~\ref{sec:ISO}).

Finally, we turn our attention to how exocometary bodies and reservoirs are affected by the evolution of the host star. Stellar winds are discussed in Section~\ref{sec:winds} and Appendix~\ref{app:winds}. Small body evolution around giant branch stars is discussed in Section~\ref{sec:GB}, and around white dwarfs (WDs) in Section \ref{sec:WD}. WDs are now known to display a host of phenomena related to cometary activity: asymmetric transits, the existence of dust on close-in orbits, and the deposition of material into the WD atmosphere. While it is unclear whether the original composition of the bodies responsible for these phenomena is volatile-rich or volatile-poor, their delivery from a source region to close to the WD is a problem closely related to the delivery of comets around main sequence stars, and we discuss them here without prejudice. We finish with a list of open questions (Section~\ref{sec:Qs}) and conclusions (Section~\ref{sec:conclude}).

\section{Getting bodies from reservoirs to stars}
\label{sec: mechanismsToGetCometsCloseToStars}

\cite{Bannister2025} detail the material reservoirs whence we think comets originate. These reservoirs are typically located in the outer regions of planetary systems, so some mechanisms are needed to drive the bodies inwards towards the star, where they can sublimate and be observed as comets or exocomets. This section describes mechanisms that can remove bodies from reservoirs and place them on high-eccentricity, low-periapsis cometary orbits. In this Section, we focus on massive analogues of our own Asteroid and Kuiper Belts, which are often visible as debris discs around other stars. Delivery from Oort Cloud analogues is discussed in Section~\ref{sec:Oort}.

\begin{figure}
    \includegraphics[width=0.99\textwidth]{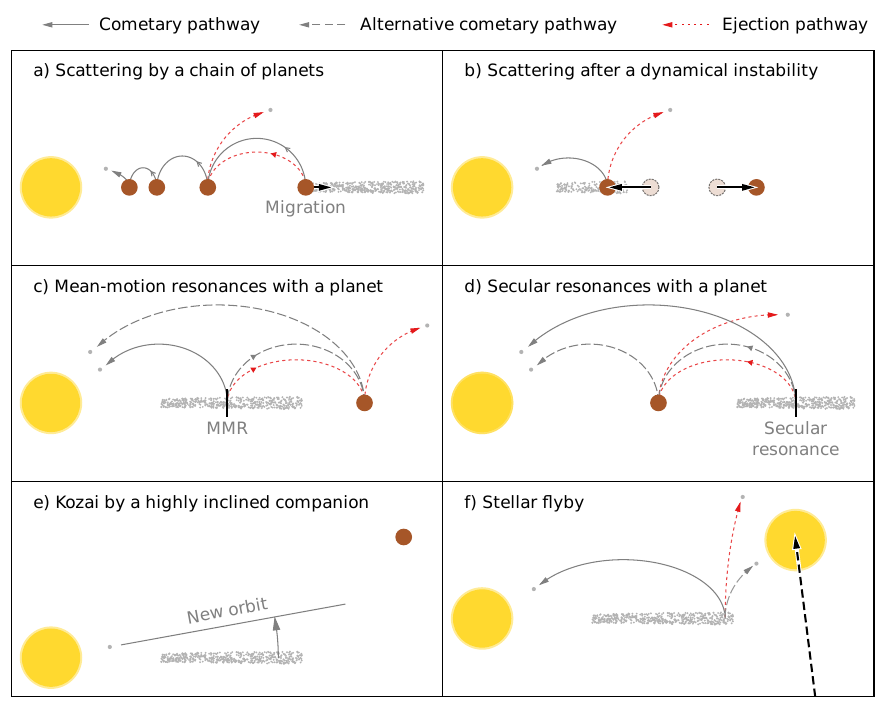}
    \caption{Possible exocomet delivery and ejection mechanisms. Cometary nuclei are thought to initially form and reside in belts or discs of planetesimals, and are subsequently either driven towards the star or ejected by various dynamical mechanisms. Each panel shows a different potential mechanism, as described in Section \ref{sec: mechanismsToGetCometsCloseToStars}. The large yellow circles are stars, the small grey points are small bodies which could become comets, and the brown circles are planets. The lines show various dynamical pathways; small bodies can either be driven towards the star to become comets (grey lines) or ejected from the system (dotted red lines). Solid and dashed grey lines show alternative pathways for getting bodies close to stars.}
    \label{fig: exocometDeliveryMechanisms}
\end{figure}

Exocomets are detected or inferred around stars of diverse ages and spectral types (\citealt{Iglesias2025} and \citealt{Korth2026ISSI}). Similarly, indirect evidence of exocomets, such as exozodical dust in the inner regions of planetary systems, is also found around a high fraction of stars with diverse ages and types (e.g. \citealt{Mennesson2014, Nunez2017, Sezestre2019, Ertel2020, Pearce2022Comets}). This implies there must be mechanisms capable of delivering exocomets at both early and late stages of stars' lives, and effective for different types of star. There is unlikely to be one universal or even dominant mechanism; a combination of mechanisms is likely responsible, with different efficiencies at different times and around different stars. \mbox{Figure \ref{fig: exocometDeliveryMechanisms}} shows some dynamical mechanisms that could create exocomets, each of which is detailed below. These are the mechanisms most discussed in the literature.


\subsection{Scattering by a chain of planets}
\label{subsec: scatteringByChainOfPlanets}

A population of small bodies orbiting close to a planet will be slowly dynamically eroded, as bodies chaotically diffuse onto planet-crossing orbits \citep{Wisdom1980,Mustill2012chaos}.
Small bodies passing near a planet will then get scattered onto significantly different orbits around the host star. A planet near a reservoir of small bodies can hence scatter them inwards towards the star, or outwards and eject them from the system. However, it is difficult for a single planet to scatter bodies all the way into the inner regions, because this would require the body to lose a significant amount of orbital angular momentum during a single scattering event. Instead, inward scattering is more easily achieved by a chain of multiple planets  (e.g. \citealt{Bonsor2014, Marino2018, Anslow2023}). In this scenario (\mbox{Figure \ref{fig: exocometDeliveryMechanisms}}a), the outer planet scatters material inwards, where it encounters the next planet. That planet scatters material inwards again, where it encounters the next planet. This process continues, and can efficiently pass bodies from an outer reservoir into the inner regions. The act of scattering material inwards also drives the outer planet outwards to conserve the system's total angular momentum, where the planet can encounter fresh material.

The efficiency of this process depends on the specific architecture of the planetary system. \cite{Rodet_2024} performed a numerical investigation of two-planet systems, and find that two planets are much more effective at producing comets than single-planet systems. They find that the efficiency by which bodies are placed onto comet-like orbits is predominantly set by the parameters of the innermost planet. Generally the inner planet is more effective at perturbing material, at least over short time-scales, because it has a shorter orbital period, which results in shorter dynamical timescales for interacting with the small bodies. 

\subsection{Scattering after a dynamical instability}

Section \ref{subsec: scatteringByChainOfPlanets} considered a chain of planets on relatively stable orbits, where the outermost planet lies near a reservoir of small bodies. However, it is also possible for planets to undergo rapid, significant orbital changes, and this can drive bodies inwards onto cometary orbits (\mbox{Figure \ref{fig: exocometDeliveryMechanisms}}b). For example, after a long period of quiescent evolution, two or more planets could undergo a dynamical instability that drastically changes the planetary orbits over a short timescale \citep[e.g.,][]{Chambers1996, Petit2020}. Such planets could gain significant eccentricities and may be driven into the reservoirs, similar to the orbital evolution of Neptune in planetary instability models of the Solar System (e.g. \citealt{Gomes2005}). This process could scatter a significant amount of small bodies both inwards and outwards, some of which could be passed inwards between planets like in \mbox{Section \ref{subsec: scatteringByChainOfPlanets}}. A late dynamical instability has been hypothesised to explain extremely dusty old debris discs such as $\eta$~Corvi \citep{Lisse2012}.

\subsection{Mean-motion resonances with a planet}
\label{subsec: mmrsWithPlanet}

Mean-motion resonances (MMRs) occur when two bodies have orbital periods that are approximately simple fractions of each other. For example, Neptune and Pluto are in the ${3:2}$ MMR, which means that Neptune completes 3 orbits of the Sun for every 2 orbits completed by Pluto. Bodies in MMRs with a planet can have their eccentricities excited to high values by the interaction, especially if the planet has undergone migration (e.g. \citealt{Wyatt2003}). Some resonances may be inherently unstable \citep{Beust1996}, or \red{they may be} destabilised by the system's global architecture \citep{Lecar2001}, forcing eccentricities still higher. These eccentricities can either be high enough that the bodies directly attain small periapsis distances (solid grey line on \mbox{Figure \ref{fig: exocometDeliveryMechanisms}\red{c}}), or they can bring the bodies close to a planet, which then scatters them inwards (dashed grey line). Since MMRs take time to excite eccentricities, this mechanism could potentially supply material to the inner regions well into a system's lifetime (e.g. \citealt{Faramaz2017, Beust2024}). The Yarkovsky effect \citep{Beekman2006}, discussed below, can also force bodies to drift into unstable resonances from stable regions of a belt at late times (this process is ongoing in the Solar System's Asteroid Belt). MMRs have been much discussed as a delivery mechanism in the reasonably well-characterised $\beta$~Pictoris system; see Lu et al (this volume).

\subsection{Secular resonances with a planet}

Section \ref{subsec: mmrsWithPlanet} described MMRs, which occur when the orbital periods of bodies are approximately simple fractions of each other. A similar situation can occur when the orbits themselves precess (i.e., revolve). Planets' orbits change on long (secular) timescales with a characteristic set of eigenfrequencies that depends on the system architecture. If the precession period of an orbit is approximately in a simple ratio with one of these eigenfrequencies, then they form a secular resonance (e.g. \citealt{Murray1999}, Chapter 7).

Planetary orbits can precess for several reasons, including the influence of other planets, or of a massive debris disc \citep{Pearce2015, Yelverton2018, Sefilian2021, Sefilian2023}. When this happens, an object near the secular resonance with that planet can have its eccentricity pumped up to high values, whilst its semimajor axis remains constant. This means that if a planet's orbit precesses, and its secular resonance lies in a debris disc, then bodies in the disc can be driven onto cometary orbits with small periapsides close to the star (\mbox{Figure \ref{fig: exocometDeliveryMechanisms}}d). Alternatively, objects could be driven onto orbits which cross that of a planet, and that planet could then directly scatter them inwards.

\subsection{Kozai interaction with a highly inclined companion}

If bodies have mutual inclinations above $\approx40^\circ$, then the von Zeipel--Kozai--Lidov mechanism can occur (Kozai interactions for short; \citealt{vonZeipel1910, Lidov1962, Kozai1962}). For systems with a debris disc and a planet or companion star on a highly inclined orbit, this mechanism can cause bodies in the disc to oscillate between low-eccentricity orbits in the disc plane and high-eccentricity orbits in an inclined plane (\mbox{Figure \ref{fig: exocometDeliveryMechanisms}}e). The semimajor axes remain constant in this interaction, so small bodies undergoing Kozai interactions can be driven onto cometary orbits with \red{periapsides} very close to the star. Note that there is some debate about the efficiency of this mechanism, because the effects of general relatively near the star can overcome the Kozai mechanism in some cases \citep{Naoz2016}.

\subsection{Stellar flybys}

An external star passing close to a planetary system can induce significant perturbations in both systems \citep{Adams2010}. This can directly drive small bodies from the outer regions of a system onto highly eccentric cometary orbits \citep{Kenyon2004,Morbidelli2004,Pfalzner2024b}. Flybys can also trigger (possibly delayed) dynamical instabilities that destabilise planets \citep{Malmberg2011}, which then scatter small bodies onto potentially cometary orbits \citep{Flammini2023}. It is also possible for bodies to be transferred \textit{between} stars during flyby encounters \citep{Kenyon2004,Morbidelli2004,Levison2010,Jilkova2016}; the resulting orbits may have very high eccentricities, so a body can be ripped from one star and become a grazing comet around the other star (grey dashed line on Figure \ref{fig: exocometDeliveryMechanisms}f). Stellar flybys could occur at any time, but they are much more likely in young systems, where stars are closer together in their birth environment \citep{Lada2003}. Within the Galactic field after a star has left its birth cluster, flybys that can directly affect a debris disc reservoir are extremely unlikely, but wider flybys that affect (exo-)Oort Cloud bodies are common, and they combine with the Galactic tide to help deliver such bodies to small periapsides \citep{Rickman2008}.

\subsection{The view from the Solar System}

Our own Solar System is the best-studied system we have, and it offers a warning against over-reliance on single dynamical mechanisms when attempting to explain exocomet delivery, even within a single system, as most of the above mechanisms are known or suspected to have affected the dynamics of small bodies in the Solar System. Oort Cloud comets were originally implanted in the Oort Cloud via planetary scattering, some possibly during a global instability of the giant planets (see next Section). They then are affected both by stellar flybys when close to aphelion, and by planetary scattering when close to perihelion. The dynamics of bodies in the Asteroid and Kuiper Belts is affected by mean motion and secular resonances with the giant planets, and those Kuiper Belt bodies destabilised by such resonances can then be scattered inwards by the chain of giant planets to become Jupiter-Family Comets. Meanwhile, a relatively close stellar flyby may have sculpted the trans-Neptunian region, raising the perihelia of bodies such as Sedna which are no longer in direct dynamical contact with Neptune \citep{Brasser2012}, and early flybys may have contributed to populating the Oort Cloud \citep{Levison2010}. Looking at extrasolar systems, it is likely that many mechanisms are at play in each system. 

\section{Depletion of cometary reservoirs}

\label{sec:depletion}

Section \ref{sec: mechanismsToGetCometsCloseToStars} described different mechanisms by which small bodies can be removed from reservoirs and placed onto cometary orbits. These mechanisms, combined with collisional erosion, would cause a cometary reservoir to deplete over time. This section summarises how this depletion occurs.

\subsection{Reservoir depletion via planetary interactions}
\label{subsec: reservoirDepletionByPlanets}

Our best example of a cometary reservoir that has been dynamically depleted is the Kuiper Belt, located beyond the orbit of Neptune in the Solar System. The Kuiper Belt's present-day mass is of order ${0.01 \; {\rm M}_\oplus}$ \citep{Fraser2014}, but its initial mass is thought to have been around ${20 \; {\rm M}_\oplus}$ \citep{Nesvorny2018NatAs}. This implies a depletion in mass by three orders of magnitude.

The Kuiper Belt's depletion is thought to have been largely the result of a planetary instability early in the Solar System's history. The Kuiper Belt contains scattered and resonant populations, which indicate that Neptune once migrated outwards into the disc \citep{Malhotra1995, Gomes2005}. The Solar System's giant planets may originally have had a much more compact configuration than today (e.g. \citealt{Tsiganis2005, Morbidelli2007AJ}), mainly due to planet migration in the protoplanetary disc (e.g. \citealt{Masset2001MNRAS, Morbidelli2007Icar, Kley2012ARA&A}). However, this compact planetary configuration would have been unstable after the gas disc disappeared, possibly following migration of Neptune outwards as it scattered TNOs inwards, ultimately leading to a dynamical instability \citep{Gomes2005, Levison2011AJ, Griveaud2024A&A}. The timing of the instability is not well constrained relative to the disappearance of the gas disc (e.g. \citealt{Deienno2017AJ, Clement2018Icar,  deSousa2020Icar}), but as a result Neptune was driven further outwards into a primordial Kuiper Belt that was at that time much more massive than today. This process would have scattered significant amounts of material, leading to the depletion of the Kuiper Belt. The current structure of the Kuiper Belt, and of the Jupiter Trojan population, can be well explained by this paradigm \citep{Nesvorny2013ApJ, Nesvorny2018NatAs}.

There may also be evidence that exoplanets scatter and deplete small body reservoirs in extrasolar systems. Some debris discs have sharp edges, which may be indicative of sculpting by unseen planets \citep{ImazBlanco2023, Pearce2024Edges}. In addition, \cite{Geiler2019} found evidence that the disc of \mbox{HR 8799} (where four giant planets have been detected) may contain a scattered population, and \cite{Matra2019} find evidence of both kinematically hot and cold\footnote{``Hot'' and ``cold'' here refers to the degree of orbital excitation rather than thermodynamic temperature, a nomenclature also used for the hot and cold populations of the Kuiper Belt, and for stellar populations in the Galactic disc.} populations in the \mbox{$\beta$ Pictoris} debris disc, which may also be due to scattering. The gaps in several debris discs, like \mbox{HD 107146}, may have been depleted via scattering from a hypothetical planet embedded in the disc \citep{Friebe2022}; alternatively, these may have been depleted by secular resonances with an internal planet \citep{Pearce2015, Sefilian2021, Sefilian2023}. It therefore appears that planets may deplete small body reservoirs in extrasolar systems too.

\subsection{Reservoir depletion via stellar flybys}

\begin{figure}
    \centering
    \includegraphics[width=\textwidth]{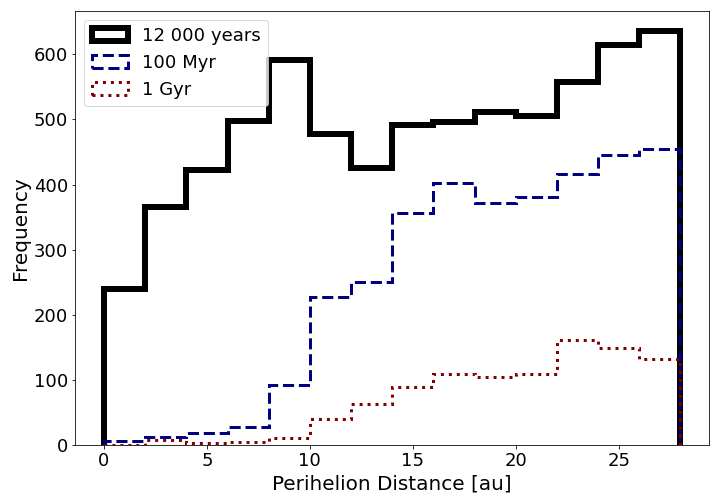}
    \caption{Depletion of the population of small bodies injected into the inner regions of the Solar System by a stellar flyby, from a simulation by \citet{Pfalzner2024}. The plot shows the number of former TNOs whose perihelion lies within each bin. The material depletes over time (measured from the perihelion of the flyby), through interactions with the giant planets.}
    \label{fig: reservoirDepletion}
\end{figure}

The extreme dynamics occurring during a stellar flyby could also significantly deplete small body reservoirs. This has been suggested in the Solar System, where a close stellar flyby potentially restructured the Kuiper Belt \citep{Pfalzner2024}. This interaction may have directly ejected a quarter of the Belt, and injected a further 9\% of the debris disc's mass into the region inside Neptune, where it would have contributed to the Centaur and short-period-comet populations. That injected population would then undergo further depletion due to interactions with planets; Figure \ref{fig: reservoirDepletion} shows the depletion of the injected small body population over time, following a flyby \citep{Pfalzner2024}. On timescales of \mbox{10 Myr} to \mbox{1 Gyr} after the flyby, the injected population is significantly depleted by the planetary interactions, particularly with Jupiter. Hence much of the material that was injected onto cometary orbits by the flyby would be quickly depleted by planetary interactions. 

\subsection{Collisions in small body reservoirs}
\label{subsec: collisionsInReservoirs}

Even without planets or flybys, a small body reservoir would naturally deplete over time. This is due to the collisional erosion of the constituent bodies. Unlike collisions in protoplanetary discs, which can be agglomerative and cause objects to grow into larger bodies, collisions in debris discs are destructive and cause objects to break apart \citep{Wyatt2008}; this is because of the lack of damping and resulting high velocity dispersion of the population of small bodies. The debris from these collisions then undergoes further collisions, breaking into even smaller objects, and this process continues until objects are ground down into dust \citep{Dohnanyi1969}. Small dust is strongly affected by radiation and/or wind from the star, so when debris breaks into small dust grains, these forces overcome gravity and the grains are blown out of the system. This continuous collisional grinding and removal of material causes debris discs to lose mass over time \citep{Wyatt2007, Lohne2008}. The short lifetime of dust grains also means that observed debris discs must be currently collisionally active.

\subsubsection{Evidence for collisions in debris discs}

Evidence for this collisional process comes from multiple sources. First, there is a clear trend between debris disc brightness and system age \citep[see][]{Bannister2025}. This means that the mass of dust in the system decreases with age, as would be expected from a collisional model. This has actually been observed in real time in the white dwarf debris disc WD~0145+234 \citep{Swan2024}, where timescales are extremely short owing to the small orbital radius of the disc. Second, observations of debris discs mainly detect dust grains with sizes similar to the observing wavelength, and the dust we see would have collisional and dynamical lifetimes much shorter than the age of the system \citep{Hughes2018, Pearce2026}. This means that this dust must be continually replenished, and collisions between larger debris bodies is the favoured means to do this. Third, IR spectroscopy of some debris discs reveals the presence of species such as SiO gas likely arising from high-velocity collisions \citep{Lisse2009}. Fourth, two transient companions to the debris disc host Fomalhaut have been detected, which are now thought to be dust clouds generated by planetesimal collisions \citep{Kalas2008,Kalas2026,Gaspar2020}. Dust release following such collisions has also been observed in the Asteroid Belt \citep{Jewitt2010,Bodewits2011}. Fifth, many Solar System asteroids can be grouped into families with similar proper orbital elements, which was early realised to indicate an origin from a common parent \citep{Hirayama1918,Nesvorny2026}. The size distributions of these families are indeed in good agreement with a collisional origin \citep[e.g.,][]{Davis1985,Marschall2022AJ}.

\subsubsection{Collisional cascade}
\label{subsec: collisionalCascade}

When material in a debris disc starts breaking up, it enters the ``collisional cascade''. This describes the process by which larger bodies undergo collisions and break into smaller bodies, which then undergo further collisions and break into even smaller bodies, and so on. In steady state, and assuming scale invariance, this leads to a well-defined size distribution of colliding bodies in the cascade. This is approximately a power law
\begin{equation}
n(s) {\rm d}s \propto s^{-p} {\rm d}s,
\label{eq: dohnanyiSizeDist}
\end{equation}
where ${n(s){\rm d}s}$ is the number of bodies with radii between $s$ and ${s+{\rm d}s}$ and $p=3.5$ for an ideal infinite collisional cascade \citep{Dohnanyi1969}\footnote{This can also be expressed in integral form, usually as a survival function $N(\ge s) \propto s^{-q}$ with $q=p-1$; this is the form shown in Figure~\ref{fig: solarSystemSizeDist} below. The diameter $D=2s$ may also be used instead of the radius.}. The steep index of $s^{-3.5}$ means that smaller bodies vastly outnumber larger ones. It also means that the total surface area (and hence thermal emission) of the disc is dominated by the smallest bodies, whilst the total mass is dominated by the largest bodies. Real discs may not exactly follow the size distribution of \mbox{Equation \ref{eq: dohnanyiSizeDist}} because the assumption of scale invariance is broken: at the larger end, the largest bodies may follow the primordial distribution, because these bodies take longer to enter the cascade (e.g. \citealt{Krivov2021}). At the smaller end, the distribution is cut off by the blowout size, which causes ``waves'' to appear in the distribution of small grains. At intermediate sizes (around 100\,m), a change in bodies' strength from being dominated by material strength to gravity induces further deviations, recently detected in observations of Main Belt Asteroids \citep{Burdanov2024}. There may be further deviations from the simple powerlaw, for example if larger bodies undergo cratering collisions that break off relatively small fragments without significantly disrupting the body (e.g. \citealt{Thebault2007, Desch:2021}). Simulations of collisions with different parameters show that the size distribution of fragments from a single collision can be convex or concave \citep{Durda2007}, and a steep power law distribution can persist in some collision families such as the Eurybates family \citep{Marschall2022AJ}. Nonetheless, \red{in collisionally evolved populations such deviations tend to relax, and then} \mbox{Equation \ref{eq: dohnanyiSizeDist}} is still a reasonable approximation of the size distribution.

Simulations of the long-term collisional evolution of populations of Solar System bodies broadly reproduce power-law distributions of sizes. 
Figure~\ref{fig: solarSystemSizeDist} shows the simulated size distribution of the main Asteroid Belt \citep{Bottke2020AJ} and the Kuiper Belt's scattered disc (``destabilized population'' \red{in Figure~\ref{fig: solarSystemSizeDist}}; \citealt{Bottke2023PSJ}), early in the Solar System's lifetime. These distributions roughly follow the simple form of \mbox{Equation \ref{eq: dohnanyiSizeDist}} at small sizes, but are shallower and wavy at larger sizes because \red{of} the dependence on changing strength at $\sim100$\,m. The size distribution of the two populations differ, because they do not follow the same disruption law. Kuiper Belt Objects (KBOs) are easier to break up than asteroids, because they have different material compositions and bulk densities; the weakest KBO body (i.e., the easiest to break up) would have a diameter of roughly 20~m, compared to 200~m for an asteroid \citep{Bottke2023PSJ}. 
\red{Using debiased NeoWISE survey results, \cite{Bauer2017} found that the size distribution of Jupiter Family Comets have a similar differential size distribution $(\propto D^{-3.3})$ to an ideal collisional cascade, while Long-Period Comets have a much flatter size distribution $(D^{-2})$, which may reflect different histories of collisional or insolation processing between the two populations.}

\begin{figure}
    \centering
    \includegraphics[width=0.99\textwidth]{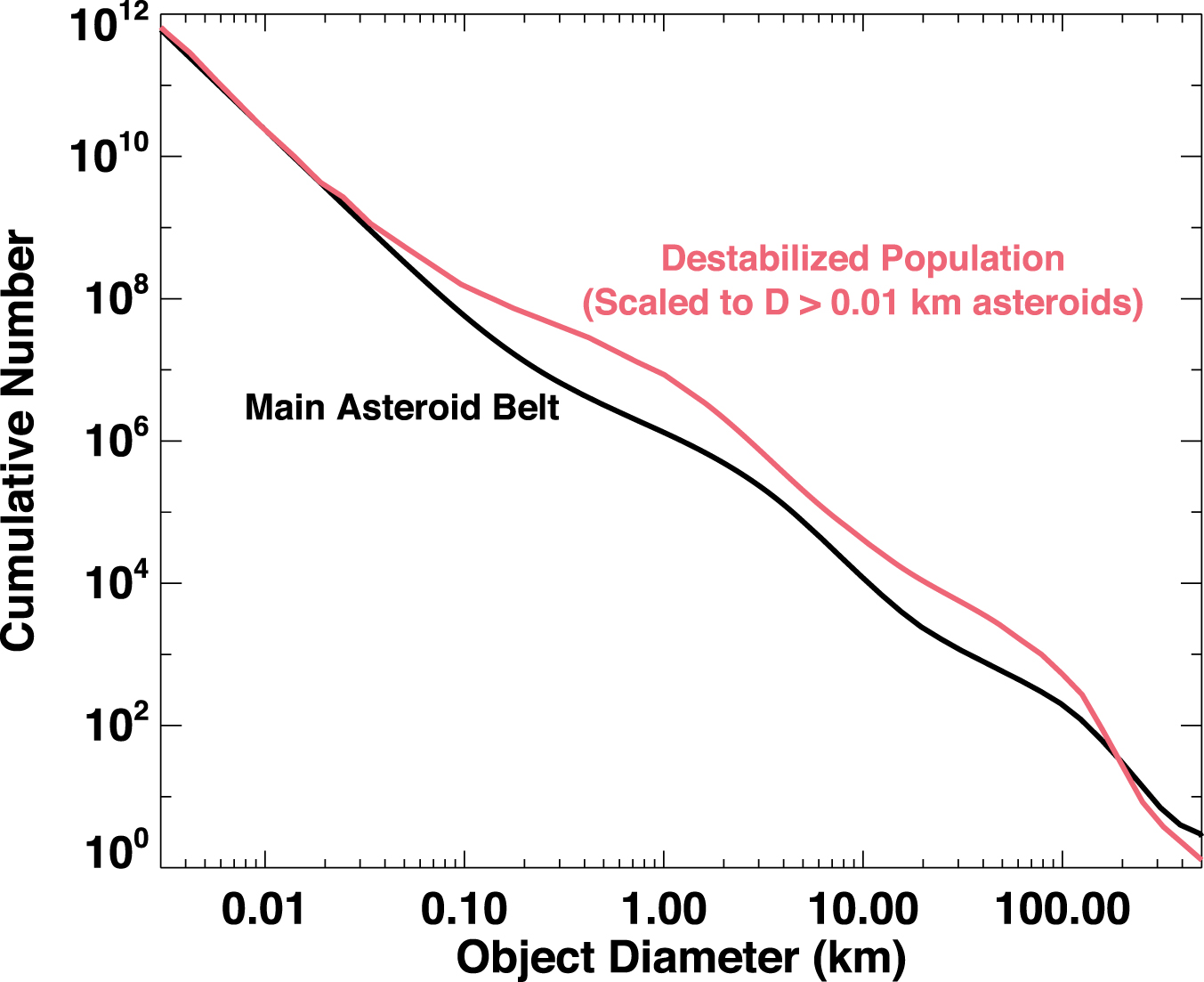}
    \caption{Simulation of the early size distribution of the main Asteroid Belt (black line, \citealt{Bottke2020AJ}) and the trans-Neptunian Scattered Disc (``destabilized population''; red line, \citealt{Bottke2023PSJ}). The Scattered Disc population is scaled to match the number of bodies with diameter above $0.01$~km. Figure from \citet{Bottke2023PSJ}, reproduced under the CC-BY-4.0 license \url{https://creativecommons.org/licenses/by/4.0/}.}
    \label{fig: solarSystemSizeDist}
\end{figure}

\subsubsection{``Stirring''}

The collisional cascade model provides significant insight into debris disc evolution, but there are still several unanswered questions. First, in order for collisions to be destructive, the bodies must have high relative velocities. This means they must have excited orbits; i.e., their eccentricities and/or inclinations must be significantly non-zero. However, when planetesimals form they are thought to have relatively unexcited orbits, because gas in protoplanetary discs should significantly damp the orbits of young planetesimals. If true, this means that debris must get excited after the protoplanetary disc phase, in a process called ``stirring'' \citep{Wyatt2008}.

We do not know the mechanism(s) behind stirring, although many have been suggested. One possible stirring mechanism is the gravity of the larger planetesimals within the disc itself, which could excite the other planetesimals \citep{Kenyon2001, Kenyon2008, Kenyon2010, Kennedy2010, Krivov2018}. However, in some cases the disc masses required to grow the large planetesimals needed within the age of the disc would be implausibly high \citep{Krivov2021, Pearce2022ISPY}. Alternatively, planets outside the disc could stir disc material through several processes, including secular interactions \citep{Mustill2009}, sweeping mean-motion resonances \citep{Friebe2022}, or by scattering material into the disc which itself excites other bodies \citep{Costa2024}. However, these processes \red{would} sometimes require planets \red{so massive} that \red{that they are ruled out by observations} \citep{Pearce2022ISPY}, and massive debris discs can also resist planetary stirring \citep{Sefilian2024}. \red{This means that planets cannot be the only stirring mechanism.} Other possibilities include ``flyby stirring'', where material is stirred by another star as it passes close to the disc, although it is unclear whether this process is common enough to stir debris discs on a population level \citep{Ida2000, Kobayashi2001, Kenyon2002}. \red{Combinations of the above mechanisms, known as ``mixed stirring'', may also occur \citep{Munoz2023,Marshall2026}.} Finally, bodies could somehow already be stirred by the end of the protoplanetary disc phase (e.g. \citealt{Walmswell2013, Booth2016, Wyatt2020}), or get stirred as the protoplanetary disc dissipates (e.g. \citealt{Zheng2017, Best2024}). However, despite considerable work into these various mechanisms, it is still unclear how debris discs are stirred.

If debris discs act as reservoirs for high-eccentricity comets, then it is possible that the same process that drives some objects onto high-eccentricity orbits also excites other objects onto collisional orbits. If true, then we might expect the source reservoirs of high-eccentricity comets to be detectable, because they should have a high level of excitement, which would lead to increased collisions and hence dust.

\subsubsection{Giant collisions}
\label{subsec: giantCollisions}

The steady state collisional cascade model in Section \ref{subsec: collisionalCascade} provides a good explanation for debris disc properties. However, in some cases, there is evidence for more extreme, stochastic collisions between larger objects. These are called ``giant collisions'' or ``giant impacts''.

Giant collisions in young planetary systems have been invoked in several scenarios to explain different observables. For instance, in our Solar System, a giant collision is the most accepted explanation for the formation of the Moon \citep{Canup2001}, and may explain the extreme morphology of Mercury \citep{Benz1988}, although Mercury's extremely high iron fraction may instead be explained by iron-rich grains in the protoplanetary disc \citep{Johansen2022}. Outside the Solar System, 
some debris discs classified as ``extreme'' have been suggested to undergo giant collisions in the past years \citep{Meng2014,Su2019}, based on large flux increases observed in their lightcurves.

Previously, the connection between giant collisions and the presence of exocomets seemed unclear, because until recently there were no systems where both occurrences were taking place. However, the advent of JWST has revealed at least two likely recent collisions in the $\beta$ Pic disc, the most prolific exocomet system. The morphological structure dubbed the ``cat's tail'' \citep{Rebollido2024} has been suggested to have originated from a possible collision point where dust is being produced sporadically. However, the distance from the star of a hypothetical collision point ($\gtrsim100$\,au) makes it unlikely that there is a direct connection between such \red{a} collision and the presence of the exocomets seen in transit. Much closer to the star is the collision claimed \red{by \cite{Chen2024}}, inwards of 10 au, who observed that the population of small grains detected two decades earlier \citep{Chen2007,Lu2022} has now disappeared, attributing this to the removal from the inner system by radiation pressure of a large quantity of debris from a large stochastic collision. Such a collision could be linked with destabilization of material in the inner regions, and therefore with a similar mechanism that is producing the exocomets. However, pinpointing the precise location of the collision, along with a dedicated dynamical model, would be necessary for an accurate determination of a link between such an event and small bodies thrown to the inner region of the system.

\section{The Oort Cloud and exo-Oort clouds as a source for comets}
\label{sec:Oort}

The Oort Cloud is a vast, distant reservoir of icy bodies surrounding the Solar System, hypothesized to explain the origins of long-period comets \citep{Oort1950}. 
\red{The Oort Cloud is traditionally divided into an inner, flattened region along the ecliptic plane (sometimes called the ``Hills cloud'' after \citealt{Hills1981}) and an outer, more isotropic spherical shell \citep{Dones2004}.} 
Its sparsely populated extent stretches from roughly 2\,000 or 3\,000 au through to approximately 0.5 parsecs ($10^5$\,au) from the Sun \citep{Fouchard2020}.
The outer fringe serves as the outermost small-body population of the Solar System, marking the region where the Sun's gravitational influence weakens against external stellar and Galactic perturbations \citep{Dones2015}. 
Most of our understanding of this distant region comes from analyzing the orbital element distributions of long-period comets. 
The structure of the Oort Cloud itself is primarily inferred from numerical simulations, which provide the basis for hypotheses about its formation and evolution \citep{Fouchard2020}.

\red{The ``Oort peak'' refers to a specific, high-density concentration in the distribution of the original orbital energies of long-period comets. This spike indicates that a vast number of these comets possess nearly parabolic, extremely large orbits with semi-major axes usually $>$ 10,000 to 20,000 au.}
The apparent inner edge of the Oort peak at approximately 10\,000 au is attributed to an observational bias, as only comets with small ($\lesssim 20$~au) perihelion distances are typically detectable \citep[e.g.][]{Hills1981}, and a semimajor axis of $\gtrsim10\,000$\,au is needed for the Galactic tide to force a body's \red{perihelion} to such a low value.
\cite{Duncan1987} developed a model describing the formation of the Oort Cloud through planetary scattering. 
Their findings suggest that the cloud should have an inner edge at approximately 3\,000 au, beyond which it transitions to an isotropic distribution. 
More recent simulations suggest that, even after 4 Gyr, the Oort Cloud is isotropic only in its outermost regions. 
\cite{Emelyanenko2013} found that for semimajor axes \( a < 6\,000 \) au, the cloud retains a strongly flattened structure, while at \( a > 8\,000 \) au, external perturbations gradually reshape it into a more isotropic distribution. 
\cite{Fouchard2018} further demonstrated that the Oort Cloud becomes fully isotropic beyond 10\,000 au, whereas its innermost region, with \( a < 2\,500 \) au, remains closely aligned with the ecliptic.  

Once emplaced in the Oort Cloud with a semimajor axis $\gtrsim10\,000$\,au, a body's orbit is affected by tidal forces from the Galactic potential and by passing stars, which can force its \red{perihelion} \red{inwards} to produce the isotropic comet population \citep{Rickman2008}. The Galactic tidal potential also causes an erosion of the Oort Cloud over Gyr, as bodies are stripped from the Sun to form tidal streams \citep{Portegies2021,Forbes2025}, which extend ahead of and behind the Solar System\red{, following similar Galactic orbits.} 
These are subsequently dynamically heated in the potential by encounters with giant molecular clouds (GMCs) and by dynamical structures such as the \red{Galactic} bar, dispersing the streams further.
GMC encounters do little to Oort clouds themselves, but are surprisingly frequent \citep{Pfalzner2020}.
A system's Oort cloud is also subject to stripping by stellar flybys, with varying likelihood depending on the star's distance from the Galactic centre.
The strength of the eroding tide also varies with the star's position in the potential as it orbits in the Galaxy. 
While the tidal force is greatest when a star crosses the midplane of the Galactic disc, it \red{also varies} if the star's Galactocentric location changes with time. This occurs for any star on a non-circular orbit, but even stars on near-circular orbits can radially migrate several kpc in their lifetimes through interactions with the Galactic bar and spiral arms, in a process called ``churning'' \citep{Sellwood2002,Minchev2010}. 
Whether this has happened through past orbital evolution for the Sun, and what it means for our own Oort cloud, is uncertain \citep[e.g.][]{Kaib2011, Minchev2013,MartinezBarbosa2015}.
While scattering by giant planets naturally gives rise to a large population of bodies on extremely eccentric orbits, Oort cloud formation is not necessarily universal for all systems; for instance, Oort clouds may have lower survival probabilities in dense cluster environments \citep{Nordlander2017} where bodies are unbound at comparatively small distances owing to frequent stellar flybys and the strong cluster tide.
In general, Oort cloud prevalence around other stars in the Galaxy is unknown.

Oort Clouds form when planets weakly scatter small bodies so that they do not attain escape velocity from the star and instead remain bound, undergoing a random walk in orbital energy (reciprocal semimajor axis), as described in \cite{Duncan1987}. As their orbits lengthen, interaction with the Galactic environment (passing stars and the Galactic tide) lifts their \red{periapsides} and temporarily traps them in the Oort Cloud at large semimajor axis. For many bodies this \red{trapping} process does not occur, and they are ejected from their host system to become interstellar objects, discussed \red{in Section~\ref{sec:ISO}.}



\section{Observationally linking comets to their source reservoirs}

\label{sec:observations}

So far this \red{work} has described the mechanisms that can generate comets from small body reservoirs, and how those reservoirs evolve over time. Since we can detect both exocomets and these reservoirs in the form of debris discs, it is pertinent to ask whether we can determine the specific sources for individual observed comets.

This is frequently attempted for Solar System comets. For example, the Kreutz family of Sun-grazing comets are best modelled as originating in the Oort Cloud and then undergoing significant orbital evolution, based on dynamical arguments \citep{Fernandez2021}. Composition arguments can also be used; for example, Comet 96P/Machholz has an unusual composition that may point towards a Kuiper Belt origin, although an interstellar origin is also possible \citep{Schleicher2008}. \red{C/2016 R2 (PanSTARRS) has high abundances of CO and N$_2$ \citep{Biver:2018,Cochran} which may indicate formation in a cold environment \citep{Mousis2021} or that it is a fragment of a Pluto-like body \citep{Biver:2018}.}

\red{However, major challenges here are that the storage and release of volatiles in comets is poorly understood and that surveys of chemical inventories are limited in the number of comets, number of species, and range of heliocentric distances over which observations are taken, though even the limited sample available to date indicates a complex taxonomy \citep{DelloRusso2016,Eistrup2019}. The Rosetta results on 67P/Churyumov--Gerasimenko show an even more complex picture with adsorption and release processes affecting observed abundances \citep{Luspay-Kuti2022,Mandt2024}. These Rosetta results have not yet been fully linked to ground based observations \citep{Saki2024}.}

For extrasolar systems, linking an exocomet to its source is more challenging. \mbox{$\beta$ Pic} is so far the only well-characterised system with frequent observations of star-grazing comets (Lu et al., this issue). The system hosts two giant planets on well-determined orbits, along with multiple small body populations at a range of distances from the star \citep{Smith1984, Lagrange2010, Matra2019, Nowak2020}. \cite{Beust2024} used dynamical arguments to show that a hypothetical reservoir interior to \mbox{$\beta$ Pic}'s inner planet could be the source, and this source could be consistent with detections of warm dust in the system. However, this is not a unique possibility, and it is also plausible that the comets originate further out in the main disc and move inwards through other processes (e.g. scattering by a chain of planets\red{; \citealt{Jaworska2026}}).

For other exocomet systems we know even less. \mbox{HD 172555} has comets detected in both spectroscopy and photometry, and hosts a debris disc that could be a comet source \citep{Smith2012, Grady2018}. However, much less is known about this system than \mbox{$\beta$ Pic}, so confidently linking the comets to the disc is difficult. Likewise, many systems with spectroscopic absorption features attributed to comets also host debris discs \red{\citep{Iglesias2025}}, but we do not have enough evidence to say that the comets came from those discs. We saw in Section~\ref{sec: mechanismsToGetCometsCloseToStars} that many of the dynamical delivery mechanisms involve planets. However, a major challenge in understanding the dynamics of exocomet systems is that the planets most favourable for driving delivery would be distant from their stars and \red{of} low mass, which makes them extremely difficult to detect. To date, very few planets have been detected in the outer regions of debris disc systems, and linking exocomets to their sources remains challenging. Furthermore, the host stars of spectroscopically-identified exocomets are more massive than the Sun \red{\citep{Korth2026ISSI}}, as are the typical progenitors of polluted WDs (Section~\ref{sec:WD}); around these stars, our knowledge of the planetary population is less than it is for FGK stars. Binary stellar companions are much easier to detect, however, and have been implicated in driving the dynamics for the photometric comet-host candidate KIC~8462852 \citep{Young2024}.

There are ongoing efforts to find the wide-orbit exoplanets that could be responsible for generating exocomets. Microlensing will soon prove fruitful on a population level, with the advent of the Nancy Grace Roman Telescope \citep{Akeson2019}, but is ill-suited for particular systems. Coronagraphic direct imaging will push down to lower masses with Roman and possibly the Habitable Worlds Observatory \citep{NAS2021,Harada2024}, though not necessarily for planets on wide orbits. Indirect dynamical traces may also prove possible in some specific systems; features seen in debris discs such as warps, eccentric belts, gaps, clumps and steep edges have all been used to infer the properties of unseen planetary perturbers (e.g. \citealt{Mouillet1997, Pearce2014, Sefilian2021, Friebe2022, Booth2023, Pearce2024Edges}). Similarly, distortions of the gas kinematics of protoplanetary discs could identify outer planets during formation \citep{Pinte2019}.

\section{Non-gravitational and tidal forces on small bodies}
\label{sec: nonGravAndTidalForces}

So far, this \red{work} has discussed the evolution of the orbits of small bodies under the influence of gravity alone. However, non-gravitational forces also play a role, and bodies can also be dynamically disrupted through tidal forces near stars and planets. This section describes the effect of these forces on small bodies. For more in-depth reviews, see for example \cite{Vokrouhlicky2015} and \cite{Jewitt2025}.

\subsection{Rotational spin-up}

The spin of a small body can be affected by several factors, one of which is the absorption and re-radiation of stellar radiation, following redistribution of the incident energy within the body. When a body is aspherical, this can generate asymmetric torques on it, altering its spin, causing the body to enter a tumbling spin state, or even accelerating it to rotational fission \citep{Paddack1969,Rubincam2000}. This physical effect is named the Yarkovsky–O'Keefe–Radzievskii–Paddack (YORP) effect\red{, reviewed by} \cite{Vokrouhlicky2015}. YORP has been directly measured through long-baseline observations of asteroids' phase curves, where a YORP torque induces a quadratic drift \red{in the timing of phase curve maxima} \citep{Lowry2007,Taylor2007}.

The spin of a cometary nucleus (or other body with significant volatile content) can also be affected by sublimation, also known as outgassing, which occurs when stellar radiation heats up the nucleus. Outgassing can occur in two broad flavours: either localised to a particular region of the nucleus, or across the entire body. The latter is similar to YORP, but driven by outgassing rather than radiation; it also leads to net torques on asymmetric bodies.  Since sublimating gas molecules carry orders of magnitude more momentum than photons, this ``SYORP'' can generate much stronger torques than YORP, and hence spin up much larger bodies or result in large changes in the rotation period in a relatively short amount of time; one example is the increase of the rotation period of comet 41P by around 30\,hr during its apparition in 2017 \citep{Bodewits2018}. Since comets have relatively low strength, sub-kilometer nuclei of small comets are likely rapidly destroyed through torque-induced rotational instability, and this effect is even more pronounced for sun-grazing comets \citep{Jewitt2021}.

YORP and outgassing are both effects of irradiation from the host star, and therefore become stronger for more luminous stars. All of these effects have been primarily studied for Solar System comets. The physics for exocomets orbiting Sun-like stars would be similar; the effects should be stronger for A-type stars, with luminosities that are \red{up to a few 10s of solar luminosities}, and stronger still for giant branch stars, with luminosities that are orders of magnitude greater than the Sun (see Section~\ref{sec:GB}).

\subsection{Orbital changes}

Similar to rotational spin-up, orbital changes of small bodies can occur due to the action on the orbit of radiative torques or/and outgassing. If the body is small enough and the incident radiation is strong enough, then these torques can noticeably shift the orbit of the body \citep{Opik1951,Beekman2006}. This physical effect is named the Yarkovsky effect, which is reviewed \red{by} \cite{Vokrouhlicky2015}. For main-sequence stars this effect is negligible on timescales well under a Myr, and certainly over the course of a single \red{periapsis} passage. In the Solar System it primarily affects bodies smaller than 30--40\,km. As with the YORP torque, this has been directly detected for some asteroids, through small deviations from their purely gravitational orbits \citep{Chesley2003}.

In the Solar System, the Yarkovsky \red{effect} acts on Main Belt asteroids to slowly move them into unstable resonances and subsequently be delivered to the terrestrial planet region to become Near Earth Objects, guaranteeing a long-term source of such objects \citep{Vokrouhlicky2015}. For comets, outgassing and perturbations from the planets may change their orbits during each perihelion passage. An often-used formulation to model the acceleration due to these changes was created by combining functional forms from \cite{Delsemme1971} and \cite{Marsden1973} into
\begin{equation}
\ \ \ \ 
\ddot{\vec{r}} = -\frac{G M_{\odot}}{r^3} \vec{r} + \mathbb{N}(r)
\left[
A_1 
\underbrace{\frac{\vec{r}}{r}}_{\begin{subarray}{c}{\rm Radial} \\ \equiv \hat{R} \end{subarray}}
+
A_2 
\underbrace{
\frac
{r \vec{v} - \vec{r} \left(\frac{\vec{r} \boldsymbol{\cdot} \vec{v}}{r} \right) }
{\left| \vec{r} \boldsymbol{\times} \vec{v} \right|}
}_{\begin{subarray}{c}{\rm Transversal} \\ \equiv \hat{T} \end{subarray}}
+
A_3 
\underbrace{
\frac
{\vec{r} \boldsymbol{\times} \vec{v} }
{\left| \vec{r} \boldsymbol{\times} \vec{v} \right|}
}
_{\begin{subarray}{c}{\rm Normal} \\ \equiv \hat{N} \end{subarray}}
\right]
,
\label{rtn}
\end{equation}
where
\begin{equation}
\mathbb{N}(r)
\equiv
\alpha \left(\frac{r}{r_0}\right)^{-\eta}
\left[1 + \left(\frac{r}{r_0}\right)^{\xi} \right]^{-\zeta},
\end{equation}
\red{\citep[e.g.,][]{Szutowicz2000,Sosa2011}.} Here, the vector $\vec{r}$ represents the distance between the centre of the comet and the centre of the star. The first term represents the star's Keplerian potential, and the second the additional acceleration due to outgassing. The function $\mathbb{N}(r)$ quantifies the strength of this perturbation as a function of distance to the star; if $\mathbb{N}(r)\equiv0$, the comet experiences no perturbations and remains on a fixed orbit. This is only possible if $\alpha=0$, where $\alpha$ represents the extent of the outgassing, or more specifically the magnitude of the sublimation rate. The constants $r_0$, $\eta$, $\xi$ and $\zeta$ are empirically derived, with typical values of \red{$r_0=2.808$\,au,} $\eta=2.15$, $\xi=5.093$, and $\zeta=4.6142$ \red{\citep{Sosa2011}}. The values of $A_1$, $A_2$ and $A_3$ are best treated as time-dependent. The radial, transversal and normal contributions to the motion are labelled in Equation~\ref{rtn}.

A general formulation such as the one above may represent a useful starting point for representing the motion of exocomets which are observed to be periodic. However, observations of individual \red{main-sequence} exo-systems have yet to definitively show periodic exocometary behaviour. Nevertheless, this formulation is general enough to model a comet's motion over any length of time, and special cases can reduce to models such as assuming that sublimation occurs only on the starlit side of an exocomet \citep{Veras2015Sublimation}.





\subsection{Disruption}

\label{sec:disruption}

\begin{figure}[h]
\centering
    \includegraphics[width = 0.38\textwidth]{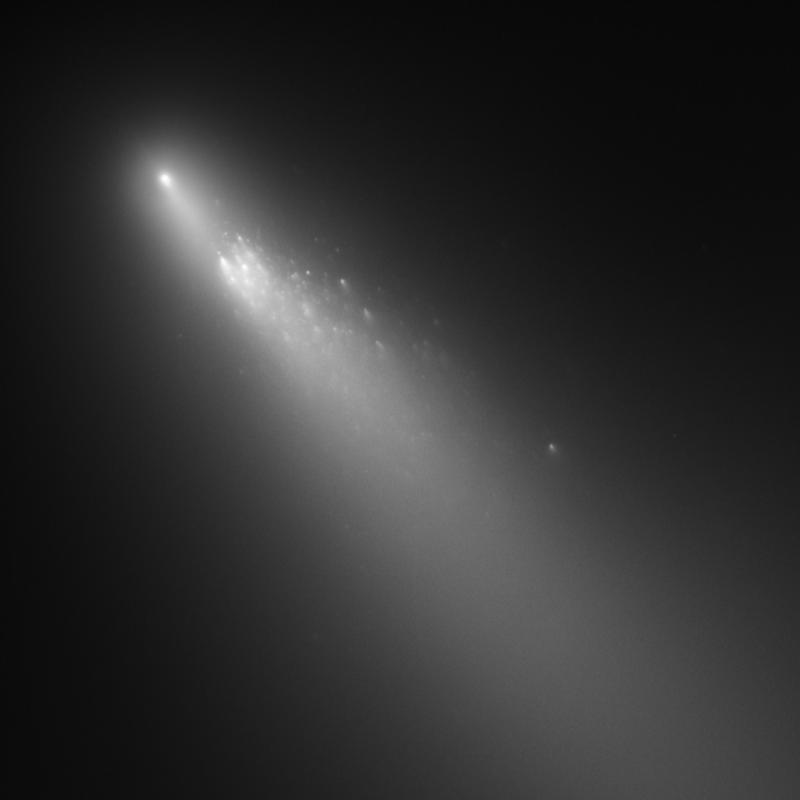}    
    \includegraphics[width = 0.58\textwidth]{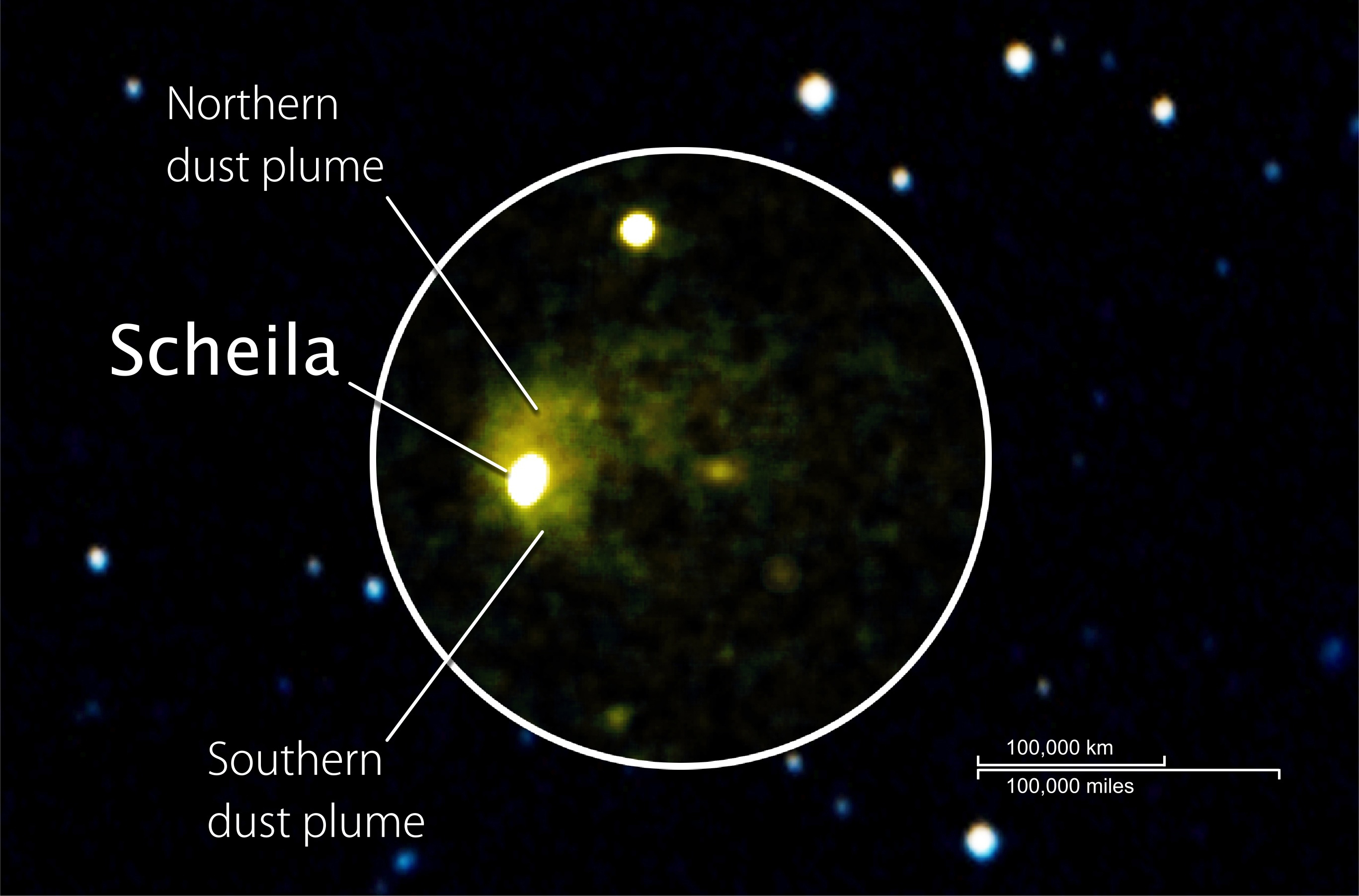}

    \caption{Disruption of Solar System bodies. Left: partial disintegration of comet 73P/Schwassmann-Wachmann 3 producing a swarm of nuclei, as seen by the \textit{Hubble Space Telescope} (NASA -- ESA -- H. Weaver: APL/JHU -- M. Mutchler and Z. Levay: STScI, \url{https://science.nasa.gov/image-detail/comet-73p/}). Right: the aftermath of an impact on asteroid (596) Scheila, as seen by the Neil Gehrels-Swift observatory \citep[][Image credit: NASA, \url{https://stsci-opo.org/STScI-01EVT4S5ZKVNHPZTD5AXSETYT1.jpg}]{Bodewits2011}.}
    \label{fig: disruption}
\end{figure}

Small bodies can disrupt partially or fully, and through a variety of mechanisms \citep{Jewitt2004, Boehnhardt2004}. The physical details of the disruption process have been described extensively for Solar System comets, with examples shown in \mbox{Figure \ref{fig: disruption}}. These processes are, however, less well-studied for extrasolar systems.

One way that objects can be disrupted is via interactions with other bodies. Total disruption is guaranteed if an object's \red{periapsis} lies within the Roche radius for tidal disruption around the star, where the object fragments as the differential tidal force across its volume exceeds its own self-gravity and material strength. The Roche radius typically lies at around $1\mathrm{\,R_\odot}$, depending on the mass and luminosity of the star, and the composition of the object. Thus, on a close approach, the object could impact the star directly, fragment due to tidal forces, or sublimate entirely \citep{Brown2011,Brown2015,Brown2017}. The mechanics of these processes are a function of the body's shape \citep{McDonald2021}. Even outside a star's Roche radius, disruption (total or partial) of one body by another can still occur. A small body could cross the Roche radius of an exoplanet (analogous to Shoemaker--Levy~9 and Jupiter; e.g. \citealt{Harris1996}) and be disrupted. Disruption is also possible somewhat outside the Roche radius, \red{as} an exchange of spin and orbital angular momentum \red{can lead} to rotational fission \citep[e.g.][]{Richardson1998,Makarov2019,Makarov2020,Veras2020ZTF}.

Several other channels exist for the disruption of small bodies. Thermal processes can destroy nuclei through complete sublimation during repeated \red{periapsis} passages; the sublimation lifetime of these comets is a strong function of stellar type, even on the main sequence \citep{Marboeuf2016}. Structural failure is another route: comets are sometimes seen to disintegrate, for example 73P/Schwassmann--Wachmann 3 (Figure~\ref{fig: disruption}, left panel). In addition, the YORP effect or anisotropic outgassing can spin a body up to centrifugal break-up speed \citep{Bottke2006, Jewitt2021}, as described above. Spin rates below the break-up rate can still lead to modest periodic mass shedding as a time-dependent function of a body's structural deformation along its orbit, as modelled for asteroid (3200) \red{Phaethon} \citep{Nakano2020}, which is not only the likely progenitor of the Geminid meteor stream \citep{Gustafson1989,Williams1993}, but also \red{has the closest perihelion of any named asteroid, at 0.14\,au.}

\red{As described above in Section~\ref{subsec: collisionsInReservoirs}, collisional processes also disrupt small bodies. In the Solar System,  collisions have been directly observed, for example the impact on asteroid (596) Scheila (e.g. \citealt{Bodewits2011}), as shown in Figure~\ref{fig: disruption} (right panel).}

To date, no direct evidence of an exocometary break-up around a main-sequence star has ever been observed\footnote{Although several disintegrating planets have been detected, such as KIC12557548b \citep{Rappaport2012} and K2-22b \citep{SanchisOjeda2015,Tusay2025}.}. As a result, active asteroids in the Solar System represent the primary observational guide for understanding how exocometary nuclei might break up. Typical examples which have prompted relevant related discussions are 311P/PANSTARRS \red{\citep{Jewitt2015episodic}} and P/2013~R3 \citep{Jewitt2017}, the latter of which spun up to break-up. A more atypical, but particularly important example, is (3200) Phaethon. With its extremely small perihelion (0.14\,au), it provides a useful testbed for the physics of the survival and destruction of small bodies close to their star \red{\citep[see][]{Vrignaud2026}}.

While we have no direct evidence of exocometary break-up, there is indirect evidence. The long duration of exocometary signatures in the \bp\ spectrum (sometimes greater than several days) indicates that those signatures may not be produced by single objects, but rather by the transit of chains of exocomets originating from the break-up of a parent body. A similar hypothesis was proposed to explain the unusual shape of a cometary transit observed in the light curve of KIC 8462852 (``Boyajian's star'', \citealt{Boyajian2016}), with at least 7 objects transiting the star within a few days \citep{Kiefer2017}. 
The unusual shape of 1I/`Oumuamua could also be \red{consistent with} tidal disruption that occurs in some cases as exocomets are ejected from their systems\red{, though erosion or evaporation from an initially much less extreme aspect ratio may offer a better explanation \citep[][and see Section \ref{sec:ISO}]{Jackson:2021}}. Returning to \bp, the ``cat's tail'' feature may be evidence of a recent collisional break-up.





\section{Evolutionary processes acting on cometary nuclei}

\label{sec:evolution}

Comets in the Solar System have been studied for decades with the expectation that their properties will reveal the conditions and physical properties during the epoch of planet formation, because they have spent most of their lives at large heliocentric distances, less affected by insolation and the solar wind than bodies closer to the Sun. 
This has motivated a series of successful space missions to comets, including \textit{Giotto}, \textit{Stardust}, \textit{Deep Impact}, and \textit{Rosetta/Philae} \citep[see][for a history]{Snodgrass2024}.
In combination with remote observations, as well as modelling and laboratory efforts, these missions have revealed that comets are in fact geological objects which have undergone complex evolution since their formation. 
Moreover, dynamical links have been established that connect comets in the Solar System with other populations: mainly their reservoirs in the Oort Cloud and trans-Neptunian region, but also more evolved populations such as Centaurs, dormant comets, Trojans, and the irregular satellites of the giant planets \citep[c.f.][]{Jewitt2024a}. 

\begin{figure}
    \centering
    \includegraphics[width=0.99\textwidth]{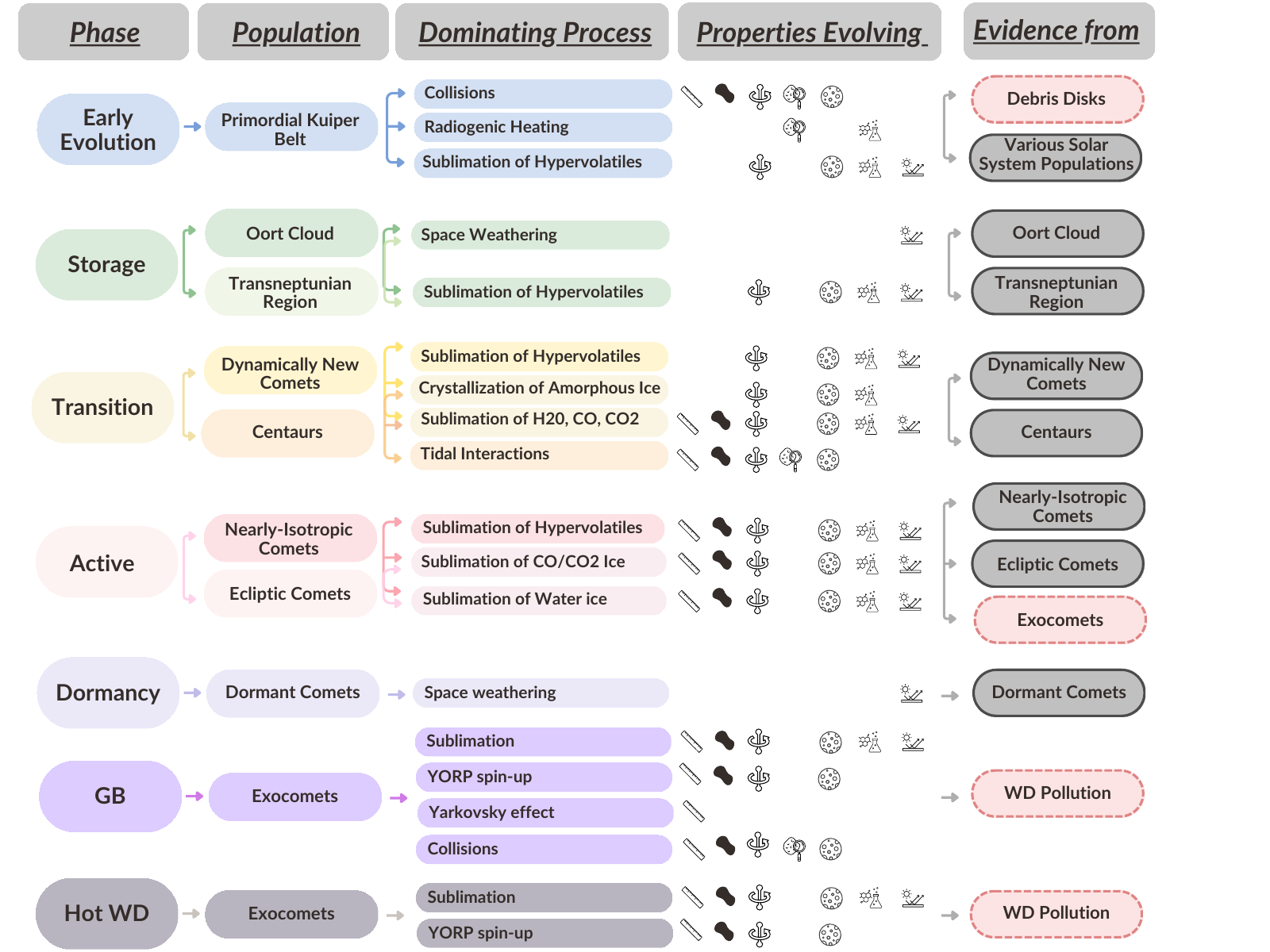}
    \caption{Schematic overview of the evolution processes acting on small bodies around solar-mass stars throughout the system's evolution\red{, from the main sequence, through the giant branch (GB), to the white dwarf (WD) phase}. The diagram
includes the most likely mechanisms driving the evolution for a given stage, and the properties which
are most likely to evolve as a result. The symbols of the properties from left to right are: size, shape, spin state,
material properties, surface topography, composition, surface reflectance. The last column indicates the observed populations which are used to extract information about the given evolutionary stage. Populations outside the Solar System are indicated in pink \red{with dashed borders in this column}.}
    \label{fig:processes-diagram}
\end{figure}



The properties of each population and the comparison between these different populations allow us to constrain the main processes believed to act on the objects. Stages of the evolution and the dominant processes identified are illustrated in Fig.~\ref{fig:processes-diagram}. 

However, by considering objects in the Solar System, we are limited to direct evidence only about the evolutionary processes taking place around a main-sequence solar-type star with a system having the architecture of the Solar System. Evidence from exocometary systems allows us to expand this study to further evolutionary stages and more diverse systems. Observations of debris discs and WD systems enable us to study the evolutionary processes taking place at early and later evolutionary stages, while the exocomets detected through photometry and spectroscopy reveal the properties of small bodies showing cometary activity around different main-sequence stars. For observations of individual cometary nuclei beyond the Solar System, we are limited at present to our \red{three confirmed} interstellar visitors 1I/`Oumuamua\red, 2I/Borisov \red{and 3I/ATLAS \citep{Meech2017,Williams17,borisov_2I_cbet,Jewitt2019,Denneau2025,Seligman2025}, discussed in Section~\ref{sec:ISO}}.

\red{A major benefit of  studying} the evolutionary processes acting on small bodies throughout their evolution is that with a better understanding of the physical processes and the properties they affect, we can link better the current properties of comets to the primordial planetesimals, and we can provide better constraints on planetesimal formation models, disc composition, mixing of small body populations, and early enrichment of the planets with organic material and water.

At \red{any} given moment, the Solar System provides \red{only} a snapshot of the current evolutionary stage of individual objects and populations. To understand the connections between different populations, we must identify similarities in object properties and patterns within populations. Following this approach, the main dynamical and genealogical links between small-body groups which were initially studied as separate populations have been identified \citep[see][]{Dones2015,Fraser2023}. Below, we provide an overview of the five stages of the comet life cycle, along with the associated populations, evolutionary processes, and properties subject to alteration. This current view on cometary evolution is summarized in Fig.~\ref{fig:processes-diagram}\red{, to which we add processes acting on comets around giant branch and white dwarf stars, relevant for other, more evolved, planetary systems}.

\subsection{Formation and early evolution}

Today's comets formed in the primordial Kuiper Belt (PKB) - a progenitor population of planetesimals beyond the orbits of the giant planets \cite[see][and references therein]{Nesvorny2018}. Within the first 100 Myr of the formation of the Solar System, the PKB underwent an extensive dispersal \cite[see Section~\ref{sec:depletion}, and][for a review]{Nesvorny2018}. This instability ejected most objects from the PKB into interstellar space. The remaining objects were emplaced into today's Oort cloud, the dynamically-excited component of the Trans-Neptunian region, irregular satellites of the giant planets, Jupiter and Neptune Trojans, and the outer Main asteroid belt. 

Assessing the early evolution of cometary progenitors remains elusive due to the scarcity of direct evidence from that epoch. Many estimates are model-dependent and are limited by the uncertainty on the planetesimal formation mechanism, location, time and timescales \red{\citep[see][]{Guilbert2023-cometsiii,Bannister2025}}. Even with these limitations, three main mechanisms are believed to have contributed at least to some extent to early cometary evolution: 1) Thermal processing due to radiogenic heating from to the decay of short-lived radiogenic isotopes (such as \textsuperscript{26}Al; see \red{\citealt{Malamud2016,Davidsson2021a, Malamud2022}}); 2) Thermal processing from solar irradiation leading to sublimation of the hypervolatile species \red{\citep{Steckloff2021,Parhi2023,Gkotsinas2024}}; and 3) collisional processing\red{, including cratering, fragmentation and catastrophic breakup,} which could have altered the physical properties of the individual objects as well as the size distribution of the whole population \citep{Morbidelli2015,Bottke2023PSJ}.


\subsection{Storage Phase}

The two main comet reservoirs are in the Oort cloud and in the dynamically excited component of the Trans-Neptunian region consisting of objects with large \red{orbital} inclinations and eccentricities \citep[see][for a review]{Dones2015}. During their prolonged residency in the outer Solar System, comet progenitors experienced minimal processing. They were mainly subject to space weathering: objects in the reservoirs experienced long-term irradiation by the solar wind, UV photons, and cosmic rays, as well as micrometeorite bombardment\red{, and potentially the interstellar medium, if at a great enough distance from the star \citep{Stern1990,Cooper2003}, as well as occasional passages with massive stars and even supernovae \citep{Stern1988}}. These processes likely affected the surface composition, as well as the colour, albedo, and the surface regolith structure \citep[e.g.,][]{Hapke2001,Cooper2003,Hudson2008,Pieters2016, Zhang2023}. Additionally, recent models show that at the heliocentric distances of the Trans-Neptunian region, \red{a combination of solar radiation and radionuclide decay would significantly deplete hypervolatiles like CO, N$_2$ and CH$_4$, though some may survive  trapped in amorphous ice \citep{Parhi2023}. At larger distances, galactic cosmic rays would also deplete hypervolatiles \citep{Desch:2021}.}


Due to their large distances from the Sun ($\gtrsim$ 2000 au), Oort Cloud objects cannot be observed directly. TNOs, however, are accessible to \red{telescopes based on or near Earth} and have even been visited by the New Horizons mission \citep{Stern2015,Stern2019}. 
Remote observations of TNOs have revealed interesting properties of the population which provide important constraints on the cometary life cycle. 
Notably, the TNO region hosts objects with a remarkable diversity of surface colours ranging from objects with neutral (solar-like) colours to objects with very red surfaces, known as ultrared objects \citep{Jewitt2002}. \red{Recent spectroscopic observations with JWST have helped to break previous degeneracies in surface composition, allowing bodies to be classified based on their surface ices and other molecules \citep{DePra2025,Pinilla-Alonso2025}.}
Another peculiar characteristic of the TNO region is the high abundance of binary and contact-binary objects, which has been used as a constraint on planetesimal formation mechanisms \citep{Nesvorny2018,Showalter2021}. 

Unfortunately, as of now, the findings about TNOs provide only indirect evidence for comet evolution. The TNOs typically observed with telescopes have radii $>$ 100 km and are therefore much larger than typical comet nuclei (1-10 km), due to the detection distances and the steep size distribution of both populations. 
The target of New Horizons, \red{(486958)} Arrokoth, is of the same size range as the larger comets; however, it is a Cold Classical Kuiper Belt Object which is believed to have formed in situ\red{.  Both Oort cloud objects and ecliptic comets are thought to have formed closer to the Sun than the Cold Classicals, in the portion of the primordial planetesimal disc subject to strong scattering from the gas and ice giant planets \citep[see][for recent reviews]{Nesvorny2018,Gladman2021}. Thus, Arrokoth may have formed further out than the comets \citep{Stern2019,Nesvorny2017}.}

\subsection{Transition Phase}

The transition phase is defined by the re-entry of an object from the reservoirs into the inner \red{regions of the Solar System where outgassing begins}. State-of-the-art dynamical simulations of the outer Solar System indicate that the orbits of some subpopulations of TNOs can be perturbed by Neptune and diffuse \red{inwards to smaller perihelia} \citep[see][]{Nesvorny2017}. Bodies undergoing this process, whose \red{perihelion} has not yet come inside Jupiter's orbit, are known as Centaurs. When encountering a planet along their way, these objects get scattered inwards and outwards in a random walk, getting passed to the next planet interior or exterior to it \citep{Duncan2004}. Centaurs begin to show comet-like activity as far as 10 au from the Sun \citep{Jewitt2009}, thought to follow recent (within a few 100 years) jumps in orbital semimajor axis \citep{Lilly2024}. At these heliocentric distances their activity cannot be driven by the sublimation of water ice as in comets, but instead can be explained either by the sublimation of \red{hypervolatiles} or by the crystallisation of amorphous ice \citep{Jewitt2009,guilbert2012}. 
Centaurs remain among the least understood objects in the Solar System: they are challenging to observe from the ground due to their faintness \red{(though a handful of detections of sublimating gas have been made, \citealt{HarringtonPinto2022,HarringtonPinto2023})} and have not yet been visited by spacecraft. \red{As with TNOs, recent JWST observations have improved our knowledge of their surfaces, which seem to be consistent with those of thermally processed bodies from the parent TNO populations \citep{Licandro2025}.}

The other population in the transition phase is that of the Dynamically New Comets (DNCs). DNCs are believed to be Oort cloud objects \red{whose perihelia are re-entering the planetary region} for the first time. \red{They are typically identified as being bodies with a semimajor axis of $\gtrsim10^4$\,au \citep{Levison1996}, although owing to the random walk in (reciprocal) semimajor axis caused by planetary encounters it is not guaranteed that an individual object at $a>10^4$\,au has never before had a small perihelion.} These objects \red{may be} the most pristine material surviving in the Solar System\footnote{\red{Note that \cite{Weissman2020} favour ``primitive'' over ``pristine'', as the former implicitly allows for some evolution since formation.}} because they have been emplaced in the Oort Cloud at large distances from the Sun, \red{and have remained in cold storage until the current epoch. The dominant process affecting cometary nuclei below their surface layer is thermal heating from \red{solar} radiation \citep{Weissman2020}, and while this is negligible once a body is ``in storage'' in the Oort Cloud, thermal processing can take place during the phase of scattering off the giant planets while the cometary nucleus is undergoing implantation into the Oort Cloud. This processing can remove volatiles such as CO and CO$_2$, and trigger the crystallisation of amorphous water ice \citep{Gkotsinas2024}: despite their supposed ``primitive'' nature, \cite{Gkotsinas2024} predict that over 1/3 of Oort Cloud objects lost all their CO early in their evolution, based on dynamical models of the early Solar System, though the average effect is still less than for KBOs \citep{Nesvorny2017,Vockrouhlicky2019,Raymond2020}. Counterintuitively, DNCs are predicted in these models to be more volatile-depleted than dynamically old comets, owing to their different early dynamical evolution. \cite{HarringtonPinto2022} indeed found such a difference, though it may also be attributed to cosmic ray processing of the surface layers \citep{Gronoff2020,Maggiolo2020,HarringtonPinto2022}.}

The upcoming mission Comet Interceptor \citep[CI;][]{Snodgrass2019,Jones2024} is uniquely designed to visit a DNC. This design, however, requires CI to be built and launched \red{potentially} before its target is discovered, and the task of identifying the most suitable \red{object{\footnote{\red{The preferred target is a dynamically new comet, or even an interstellar object, but if a suitable target is not found another long-period comet will be selected.}}}} to accomplish the \red{mission's} goals is still pending \red{by design \citep{Snodgrass2026}. Comet Interceptor may even be able to visit an Interstellar Object, but a future, more capable, mission would offer greater possibilities here \citep{Snodgrass2025WhitePaper}: ISOs are more challenging than bound comets owing to their higher velocity.}

\subsection{Active phase}

If the orbits of Centaurs evolve to cross Jupiter’s orbit, they become dynamically dominated by Jupiter and \red{(if they are not ejected)} typically have orbital periods reduced to $<20$ years, becoming Jupiter-Family Comets (JFCs). These objects get increasingly warm as they approach the \red{Sun} and are expected to begin sublimating water ice \red{around} 3--5 au. Active JFCs are significantly better understood than comets at any other stage of evolution. This is largely due to the spacecraft missions which have provided in-situ studies of six JFC nuclei \citep[see][]{Snodgrass2023}. In combination with \red{Earth-based} observations, they have allowed us to characterize JFCs in terms of their shapes, rotations, physical properties (density, porosity, strength), compositions, surface morphologies, surface properties and activity mechanisms.
In particular, Rosetta’s continuous observations of the nucleus of 67P/Churyumov--Gerasimenko throughout its perihelion provided an amazing leap in revealing the links between the nucleus properties and activity \citep{Keller2015,Keller2015a}. It led to a breakthrough in our understanding of the effects that sublimation of near-surface ices has on comet erosion \citep[\red{e.g.,}][]{Vincent2016a,Vincent2016}. Comparing Rosetta data to \red{observations from other} missions, \cite{Vincent2017} found evidence that dynamically young JFCs have surfaces dominated by steep cliffs and pits, while more evolved comets are characterised by smooth terrains covered by pebbles and dust. These differences in morphology can also be linked to different photometric properties of comet nuclei from spacecraft observations \citep{Longobardo2017,Vincent2019} and ground photometric observations \citep{Kokotanekova2017,Kokotanekova2018}. According to this hypothesis, young JFCs can be distinguished from evolved comet nuclei as they have higher albedos\red{, and their phase functions have a steeper gradient with the phase angle $\alpha$}. LPCs and DNCs entering the inner Solar System are believed to undergo similar evolution in the water-sublimation region. The main difference compared to JFCs is that, despite their larger orbital distances, they still show activity out to distances of up to a few tens of au \citep{Jewitt2021}\red{; in contrast, typical JFC aphelion distances are ${\lesssim 15 \; \rm au}$ (e.g. \citealt{Shober2024})}. 

Comets that pass sufficiently close to the Sun lose parts of their outer layers, exposing fresher, relatively unprocessed material beneath, as seen with comet 67P \citep{BockeleeMorvan2016}. Additionally, a comet’s evolution is influenced by its rotational state. For instance, \red{the high obliquity of comet 67P results in} pronounced seasonal effects that result in uneven surface evolution \citep{BockeleeMorvan2016}, \red{with northern and southern hemispheres being preferentially illuminated during different parts of its orbit as a result of the orientation of its spin axis. On the other hand,} the complex rotation of comet \red{103P/Hartley} allows solar illumination over its entire surface \red{at perihelion \citep{AHearn2011,Guilbert-Lepoutre2023}}. A \red{significant fraction} of JFCs  shows activity near aphelion \citep[e.g.,][]{Kelley2013}, indicating that regions deep within the nucleus may remain warm enough for volatile sublimation. \red{Rosetta’s observations of 67P revealed that JFC activity  is a complex interplay of near surface sublimation and the recondensation of volatiles migrating upward from deeper layers. This process can obscure the true bulk abundances  of highly volatile species such as O$_2$ \citep{Luspay-Kuti2022}.}


At sufficiently low cometary activity levels, the Rosetta mission found that the solar wind can reach the surface where it sputters refractory atoms such as  Na, K, Si, and Ca \citep{Wurz2015}. However, the interaction between the solar wind and dust in the coma has not been extensively studied \citep{Horanyi2024}\red{, though it does have an observable effect on dust tails \citep{Price2019}}. For airless bodies like asteroids, the Moon, and Mercury, exposure to the solar wind alters the properties of their surfaces through processes such as space weathering, implantation of solar wind ions, and the formation of molecules like hydrogen (H\textsubscript{2}) and water (H\textsubscript{2}O). This exposure can also result in the release of gases---including Na, K, S, and Ca---through processes \red{such as} electron-stimulated desorption and sputtering \citep{Chapman1996,Nagao2011,Wurz2022}.

Cometary comae, along with their formation and interaction with the stellar wind, are discussed further in \cite{Vrignaud2026}.

\subsection{\red{Dormancy or destruction}}

There are a few known mechanisms which determine the physical lifetime of comets. \red{Most} comets have their orbits perturbed by gravitational forces and are either ejected from the Solar System or driven onto \red{orbits with small perihelion distances} where they are eventually destroyed. Those on more stable orbits are believed to either gradually lose their activity until they become dormant comets, or, alternatively, to experience catastrophic comet-splitting events \citep[see][and see Section~\ref{sec:disruption}]{Jewitt2004,Boehnhardt2004}. One of the possible mechanisms leading to comet splitting is activity-driven spin-up. This mechanism takes place when outgassing produces torques which bring the rotation rates of the nuclei up to a critical limit. Above this limit, the centrifugal force exceeds the nucleus' gravity and the material strength, and the comet nucleus starts to shed mass and disintegrates \citep[e.g.][]{Davidsson1999}. \red{The accompanying fragmentation substantially increases the surface-area-to-volume ratio, accelerating the depletion of remaining volatiles and often producing debris streams and swarms of fragments \citep{Farnham2001, Combi2014}.}

If \red{cometary} nuclei do not undergo significant mass-loss and disruption events during the \red{peak} of their activity, \red{their evolution is instead governed by the progressive depletion or sequestration of volatiles. Continued sublimation can lead to the formation of an insulating mantle, as devolatilized material accumulates to a thickness exceeding the thermal skin depth. This mantling thermally isolates the interior, effectively sealing substantial volatile reservoirs while rendering the nucleus observationally inactive. Such objects are classified as dormant comets, whose volatiles remain present but shielded from solar insolation. With continued thermal processing, some nuclei may ultimately become dead comets, having lost their accessible volatiles entirely while retaining structural integrity \citep{Jewitt2004}. Both dormant and dead comets are observationally indistinguishable from asteroids placed on comet-like orbits \citep[see][]{Fernandez2001}. Dynamical and physical surveys estimate that dormant comets may constitute on the order of a percent of the near-Earth object population, with size-limited estimates suggesting that 0.3–3.3 \% of near-Earth asteroids with diameters larger than 1 km are dormant short-period comets \citep{Mommert2015}. Discriminating between these intact end states, and characterizing their physical differences, remains challenging due to the small sizes of cometary nuclei, their variable observing geometries, and their extremely low or absent activity \citep{Jewitt2024a}.}

\red{These evolutionary outcomes give rise to a variety of dynamically and physically distinct small-body populations \citep{Jewitt2024a}. Asteroids in Cometary Orbits (ACOs) are objects on comet-like orbits (typically low Tisserand parameter with respect to Jupiter) that show no detectable coma or tail; they are interpreted as a mix of dormant/extinct comets and asteroidal interlopers \citep{Fernandez2001}. Damocloids inhabit long-period or Halley-type orbits and are generally regarded as inactive cometary nuclei that have lost observable activity \citep{Jewitt2005, Jewitt2015active}. Manx comets are bodies on orbits like those of Oort Cloud comets that exhibit little or no activity but have asteroidal (rocky) spectral characteristics, indicating inner Solar System origins followed by emplacement in distant reservoirs \citep{Meech2016, Knight2017}. In the inner Solar System, main belt comets reside on dynamically stable asteroid-like orbits yet exhibit transient activity due to mechanisms such as sublimation, impacts, or rotational disruption \citep{Hsieh2006}; when measured, their CO$_2$ to H$_2$O ratios are extremely depleted compared to classical comets \citep{Kelley2023}. Dark comets are morphologically inactive near-Earth objects that nonetheless show non-gravitational accelerations consistent with low levels of outgassing, implying volatile release below traditional detection thresholds \citep{Micheli2023, Hui2017}. These populations illustrate the continuum between comets and asteroids, modulated by dynamical history and volatile content.}

\subsection{Asteroid evolution}

For most exocomets, it is unknown whether their nuclei are volatile-rich, as is the case for typical Solar System comets, or volatile-poor. We therefore now turn our attention to the Solar System's asteroids, as they may provide better analogues than Solar System comets for some bodies, for example the transiting exocomets of $\beta$~Pic. 
The majority of asteroids in the Solar System today reside in the main asteroid belt, a region between the orbits of Mars and Jupiter. 
This belt contains millions of small rocky bodies orbiting the Sun, representing a dynamically evolving remnant of the planetesimals that formed within the early solar nebula. 
The total mass of the asteroid belt is estimated at approximately $1.8 \times 10^{-9} \mathrm{M}_{\odot} \, (\sim 6 \times 10^{-4} \mathrm{M}_{\oplus})$, inferred by analyzing planetary perturbations from 300 major asteroids and a ring of mass for the smaller asteroids, combined with statistical extrapolation of asteroid sizes \citep{Krasinsky2002}. 
This mass is significantly lower than the estimated primordial mass of the asteroid belt, which, based on interpolated mass densities needed to form the terrestrial planets and Jupiter’s core, was on the order of 1 $M_{\oplus}$ \citep{Weidenschilling1977, Morbidelli2009}. 
This discrepancy implies that the primordial asteroid belt underwent substantial mass depletion and dynamical excitation, driven by gravitational perturbations and collisional processes. 
\cite{Morbidelli2015b} divide asteroid evolution into three major phases: 
(1) the early stage during the existence of the gaseous proto-planetary disc, when giant planets formed and migrated, 
(2) the period following gas dissipation, during which the asteroid belt experienced additional dynamical perturbations from planetary instabilities, and 
(3) the long-term evolution influenced by collisional fragmentation \citep{Bottke2023PSJ} and resonant interactions with planetary orbits \citep{Lecar1997, Nagasawa2002}. 

Repeated collisions between asteroids lead to fragmentation, generating smaller debris, reshaping surface properties through regolith redistribution and space weathering \citep{Hapke2001, Brunetto2015}, and modifying the size distribution of the asteroid population \citep{Bottke2015}. 
These collisions also contribute to the formation of asteroid families: groups of genetically related bodies that share similar orbital characteristics \citep[and references therein]{Nesvorny2015,Novakovic2022}. 
In some cases, particularly energetic impacts can excavate subsurface material, providing unique information about an asteroid's internal structure and composition. 
A notable example is (596) Scheila, which underwent a collision in December 2010, temporarily displaying comet-like activity \citep[][and see Figure~\ref{fig: disruption}]{Larson2010,Ishiguro2011}. 
Follow-up observations indicated that the impact exposed fresh, unweathered material from the subsurface, temporarily altering Scheila's spectral characteristics \citep{Bodewits2014Scheila,Hasegawa2022}.

\red{\section{Interstellar objects}\label{sec:ISO}}

Interstellar objects (ISOs) form a bridge between Solar System small bodies and exocomets: they formed in a different planetary system, but have been observed in our own Solar System on hyperbolic orbits with respect to the Sun \citep{Seligman2022}\footnote{Here we refer to bodies whose orbital eccentricities are much greater than unity: a number of Solar System comets have been observed with slightly hyperbolic osculating orbits owing to measurement error, or the effects of planetary perturbations and non-gravitational forces \red{around perihelion}. A fraction of these may actually leave the Solar System.}. 
\red{While probably abundant in the interstellar medium, they would rarely pass through planetary systems. As such, they likely constitute only a small fraction, if any, of the observed exocomets. Nevertheless, they afford us the only opportunity we realistically have to study extrasolar small bodies up close.}

The first such interstellar objects were detected relatively recently: 1I/`Oumuamua in 2017, 2I/Borisov in 2019\red{, and 3I/ATLAS in 2025 \citep{Meech2017,Williams17,borisov_2I_cbet,Jewitt2019,Denneau2025,Seligman2025}. 1I/`Oumuamua and 3I/ATLAS were discovered by surveys searching for Solar System objects (Pan-STARRS [\citealt{Chambers2016}] and ATLAS [\citealt{Tonry2018}] respectively), while 2I/Borisov was discovered by an amateur astronomer \citep{borisov_2I_cbet}. In the near future, we can expect the Legacy Survey of Space and Time (LSST), at the Vera C.~Rubin Observatory, to produce several dozen more detections \citep{Moro2009,Cook2016,Engelhardt2017,Ivezic2019,Hoover2022,Marceta2023b,Dorsey2025,Forbes2025}, enabling statistical characterisation of their properties such as compositions and kinematics.}

Like exocomets, ISOs can be used to study planetesimal formation in a range of environments \citep{Raymond2018}. 
Individual ISOs can be studied in far more depth than individual exocometary events, with detailed observations possible over a large range of \red{wavelengths and} heliocentric distances \citep{'OumuamuaISSITeam_2019, Bolin2020}.
In the future, there is also the possibility of visiting ISOs \red{\citep{Abell2024}} with missions such as Comet Interceptor \red{\citep{Jones2024,Yaginuma2025,Sanchez_2025}}. 

\red{\subsection{Observations}}

\red{Here we summarise the observations of the three hitherto-detected ISOs, and their implications. Detailed reviews of 1I/`Oumuamua and 2I/Borisov (prior to the discovery of 3I/ATLAS) can be found in \cite{'OumuamuaISSITeam_2019,Seligman2022,Jewitt2023ARAA,Jewitt2024, Fitzsimmons2024}. We give somewhat more details about 3I/ATLAS, as at present there is no review of this object. At the time of writing, 3I/ATLAS is still under observation.}

\red{\subsubsection{1I/`Oumuamua}}

\red{1I/`Oumuamua was detected when it was already on its way out of the Solar System, following its close perihelion passage interior to the orbit of Mercury at ${0.26 \; \rm au}$\footnote{\red{Orbital elements for these and other bodies are available at the} JPL Small-Body Database: \url{https://ssd.jpl.nasa.gov/tools/sbdb\_lookup.html\#/}}. It was relatively small upon detection, at most a few hundred metres in extent \citep{Trilling2018,Jackson:2021}.} High-amplitude variation in its lightcurve suggested a shape then considered unusual\red{: the best-fit shape was} a flattened disc \red{with a 6:6:1 aspect ratio} \citep{Mashchenko2019}, similar to the TNO (486958) Arrokoth \citep{Spencer2020} but more extreme (6:1 aspect ratio compared to 2:1). \red{It had a red reflection spectrum, similar to the D-type asteroids of the outer Main Belt and Jupiter Trojan regions \citep{Fitzsimmons2018}.}

The nature of 1I/`Oumuamua \red{initially caused an extensive discussion} \citep{'OumuamuaISSITeam_2019}, as it lacked detected activity but showed a non-gravitational acceleration \citep{Micheli2018,Trilling2018,'OumuamuaISSITeam_2019,Seligman2019}. \red{Non-gravitational acceleration of ISOs implies the presence of volatiles, which may inform our understanding of where in their disc they formed \citep{Seligman2022PSJ}. Given the lack of detected gas, however, the volatile species responsible for 1I/`Oumuamua's acceleration has been debated. One possibility would be that 1I/`Oumuamua was a collisional fragment rich in N$_2$ ice from a parent body like Pluto, which well reproduces the observed non-gravitational acceleration and the lack of observed outgassing or released dust \citep{Jackson:2021}, a scenario also proposed by \cite{Biver:2018} for the Solar System comet C/2016 R2 (PanSTARRS). H$_2$ ice has also been proposed for 1I/`Oumuamua \citep{Seligman2020,Bergner2023}, although it is physically challenging to form and neither Lyman $\alpha$ emission nor water (a potential progenitor) were detected \citep{Trilling2018,Sabat2023}. These unusual compositions and ISOs' long travel times through the interstellar medium have motivated several studies of space weathering from Galactic cosmic rays on these bodies \citep{Fitzsimmons2018,Seligman2020,Jackson:2021,Bergner2023}. It is likely that 1I/'Oumuamua decreased in size due to sublimation as it passed perihelion, and especially with an N$_2$ composition it may also have shrunk during its journey through interstellar space due to Galactic cosmic rays, and such erosion can explain its unusual aspect ratio \citep{Jackson:2021, Desch:2021}.}

\red{\citep{Desch:2021} predict that collisional fragments of differentiated bodies should dominate ISOs in the size range $50\mathrm{\,m}\lesssim D \lesssim 1\mathrm{,km}$. The two ISOs discovered after 1I/`Oumuamua have both more clearly resembled Solar System comets, but these are considerably easier to detect as their larger size and outgassing vastly increases their magnitude. Hopefully, the advent of the Vera C.\ Rubin Observatory's Legacy Survey of Space and Time, and its anticipated ISO discoveries, will allow such origin hypotheses to be tested.}


\red{\subsubsection{2I/Borisov}}

\red{2I/Borisov was detected on its inward approach, before it passed perihelion at ${2.0 \; \rm au}$. It passed just outside the orbit of Mars, and like 1I/`Oumuamua it exhibited non-gravitational acceleration, most likely caused by outgassing as it warmed on approach to the Sun \citep{Hui2020}. In contrast to 1I/`Oumuamua,} the brightness and observational geometry conditions of 2I/Borisov allowed for the characterization of its activity and the direct detection of multiple volatiles, including CN, C$_2$, OH, and CO \citep{Fitzsimmons2019, Bodewits2020, Cordiner2020}. \red{2I/Borisov was also sub-km in size \citep{Jewitt2017, Jewitt2019, Hui2020}. It underwent a large outburst at a heliocentric distance of 2.8\,au after its perihelion passage, which ejected a small fraction ($\sim10^{-4}$) of its mass \citep{Jewitt2020}.}
Pre-outburst \textit{HST} observations of 2I/Borisov did not detect significant lightcurve variation \citep{Bolin2020}. 

\red{The composition of 2I/Borisov was broadly similar in some respects to Solar System comets \citep{Opitom2021}: its dust had a comparable  redness, and its gas production rates showed a low C$_2$  to CN ratio,  marking it as carbon-depleted \citep{Opitom2019, AHearn1995}. It also had Ni and Fe abundances similar to Solar System comets \citep{Guzik2021, manfroid2021}.
However, 2I/Borisov was chemically distinct in other ways.} Its abundance of CO relative to water and HCN was exceptionally high \citep{Cordiner2020,Bodewits2020}, demonstrating that ISOs are not exactly \red{analogues of} Solar System comets. \red{The anomalously high hypervolatile ratio could be due to 2I/Borisov forming at a large distance from its host star \citep{Cordiner2020}, or alternatively forming around an M dwarf \citep{Bodewits2020}. These hypotheses were further explored and compared to Solar System comets in \cite{Seligman2022PSJ}.}

\red{\subsubsection{3I/ATLAS}}

\red{3I/ATLAS, with a perihelion of 1.36\,au, had a significantly more eccentric orbit than the first two known ISOs ($e=6.1$); we discuss the implications of this in terms of its Galactic origin in Section~\ref{sec:ISOkinematics}. Early images of 3I/ATLAS already showed an extended source, indicating the existence of a dusty coma at a heliocentric distance of $\sim4.5$\,au \citep{Alarcon2025,Belyakov2025,Frincke2025,Hoogendam2025,Jewitt2025,Kareta2025,Opitom2025,Seligman2025}. Precovery with Rubin Observatory Science Validation images \citep{Chandler2025} and the \textit{Transiting Exoplanet Survey Satellite} (TESS) \citep{Farnham2025,Feinstein2025,Martinez-Palomera2025} show that this activity had started months before the discovery in July, as early as May 2025 at a distance of 6.4\,au. There were other precovery observations by the Zwicky Transient Facility \citep{Ye2025}. HST observations later revealed a nucleus diameter of 1.2\,km \citep{Hui2026}.}

\red{ Pre-perihelion observations of 3I/ATLAS revealed the presence of CO$_2$, CO, OCS, and an extended source of water, likely icy grains \citep{Cordiner2025, Xing2025}. 3I/ATLAS has a reddened colour most similar to D-type asteroids, and consistent with that of 1I/'Oumuamua and 2I/Borisov \citep{Seligman2025}. 3I/ATLAS exhibits  properties that are generally consistent with Solar System comets at a similar distance, albeit with a slightly redder colour.}

\red{Like 2I/Borisov, its apparition was highly favourable for detailed characterization of its activity and composition. Pre-perihelion JWST observations at 3.3 au from the Sun revealed that the object's activity was driven by CO$_2$ and the presence of H$_2$O, CO, OCS, and water ice \citep{Cordiner2025}. Contemporaneous observations with SPHEREx corroborated these observations, and also reported an extended CO$_2$ coma and water ice absorption \cite{Lisse2025}. Its CO$_2$/H$_2$O abundance is far higher than that of Solar System comets \citep[e.g.][]{HarringtonPinto2022}, even after accounting for the low sublimation efficiency of water at such large heliocentric distances.  Water  production was also detected/inferred early \citep{Yang2025,Xing2025}. The elevation of hypervolatiles has been hypothesized to be caused by ice composition alteration via processing of the surface and near surface of 3I/ATLAS during its travel through the interstellar medium  \citep{Maggiolo2026}. Surprisingly, Ni was among the first species detected, appearing even before CN and HCN \citep{Roth2025,SalazarManzano2025,Coulson2026,Lazzarin2026,Paek2026}, which is typically the earliest species identified in distant comets \citep{Hutsemekers2026, Rahatgaonkar2025}. Near-perihelion observations revealed that the object has brightened post perihelion \cite{Zhang2026}, presumably due to increased H$_2$O activity. 3I/ATLAS had an extreme negative polarization signal in its dust coma \citep{Gray2025}. }

\red{Overall, and as with 2I/Borisov, the third known interstellar comet displayed a composition that was also consistent with formation at large stellocentric distance at cold ambient temperatures. Thermal modelling of the nucleus indicated that hypervolatiles were likely driving the activity initially observed at large heliocentric distances \citep{Yaginuma2026}. }

\red{\subsection{Ejection}}

In terms of their formation, there may be little difference between ISOs and observable exocomets, other than that ISOs were ejected from their systems, whilst exocomets got driven inwards towards their host stars. 
The dynamical mechanisms that can form ISOs \red{closely overlap with} those that can form exocomets, and in many cases a single process may form both comets and ISOs at the same time (\mbox{Figure \ref{fig: exocometDeliveryMechanisms}\red{a-d, f}}). The mechanisms that can generate ISOs are those which change a body's energy (or equivalently, semimajor axis); these include stellar flybys \citep{Pfalzner:2021}, and scattering by a planet \citep{Brasser2006}. Mechanisms that approximately conserve a body's energy (semimajor axis), such as MMRs and secular interactions, are unable to directly generate ISOs; however, they can indirectly generate ISOs by exciting small bodies onto planet-crossing orbits, so that those bodies are then scattered and ejected by planets (\mbox{Figure \ref{fig: exocometDeliveryMechanisms}\red{c-e}}).

If a planet undergoes migration driven by \red{its} scattering of planetesimals, an outward-migrating planet would scatter planetesimals inwards where they could become exocomets (although this likely requires more interior planets to further reduce the \red{periapsis} distances, \citealt{Bonsor2014}), whilst an inward-migrating planet would scatter planetesimals outward where they could become ISOs. If just one planet were involved, then this could mean that ISOs are more typically sampled from debris discs closer to their central stars (because the planet would need to be exterior to the disc and migrating inwards), whilst exocomets would be typically sampled from debris discs further away (with the planet interior and migrating outwards). This could lead to a detectable difference between ISO and exocomet populations, which may not be the case if other mechanisms dominate instead. However, multiple planets would complicate this picture; for example, an outward-migrating planet could scatter bodies inwards, where they are then ejected by an inner planet (as is thought to have happened during the migration of Neptune; \citealt{Fernandez1984,Malhotra1993}). \red{This then has implications for the volatile content of the ISOs, as they can be heated during the time when their periapsis has been reduced around the host star, possibly to a greater extent than Oort Cloud comets \citep{Gkotsinas2024}.}

One mechanism that can produce ISOs \red{considerably more efficiently than} exocomets is post-main-sequence mass loss of a star (see Section~\ref{sec:GB}), which causes loosely-bound bodies in Oort Cloud analogues to become unbound \citep{Veras2011, Hansen2017, MoroMartin2019}. Hence, some ISOs are probably produced via the post-main sequence ejection of a large portion of exo-Oort clouds. These ISOs could have fundamentally different properties and formation histories from ISOs that are ejected from an inner planetary system by other mechanisms, but they are predicted to form a minority of ISOs overall \citep{Veras2020Cluster, Levine2023}.

\red{\subsection{Galactic kinematics and chemodynamics}\label{sec:ISOkinematics}}

\red{The Galactic kinematics of ISOs  provide insights into their approximate age and original stellar population  \citep{Hallatt2020,Hopkins2025predict}. Therefore, measuring their incoming kinematics and chemical compositions can provide information regarding the formation of small bodies across time in the Galaxy \citep{Taylor2025}. This is because of correlations between age, composition, and Galactic kinematics of their stars in the Milky Way, which may be passed on to their ISO progeny.}

\red{For stellar populations, a correlation exists between age and velocity dispersion \citep[\emph{e.g.,}][]{Wielen1977,Nordstrom_2004}; this partly reflects kinematic heating from large structures such as giant molecular clouds and spiral arms \citep{Sellwood2002,Gustafsson_2016}, but may also in part reflect more violent heating for the older populations \citep[\emph{e.g.,}][]{Miglio2021}, or early star formation in a thicker, turbulent gas disc \citep[e.g.,][]{Renaud2021}.
ISOs should experience the same degree of kinematic heating as stars, as they tend to be ejected with very low velocities within a few km/s of their host, are subject to the same dynamical forcing in the Galaxy \citep{Hands2019,Pfalzner2021ISOs,Hopkins2025predict}, and so the dispersion of their velocity distribution should also increase with age \citep{Seligman2018,Forbes2025}.}

\red{Spectroscopic surveys of Milky Way stars have shown that their composition varies, with large scatter, with age \citep{Marsakov_2011}. While this age--metallicity relation has been fairly flat for the past $\sim6$\,Gyr \citep[\emph{e.g.,}][]{Haywood2013}, older, more dynamically-excited stars tend to have lower overall metallicities but higher abundances of alpha elements\footnote{\red{Alpha elements are those formed by successive additions of helium nuclei, such as oxygen, magnesium, etc.}} relative to iron \citep[\emph{e.g.,}][]{Edvardsson1993,Bensby2003,Bovy_2012VerticalMotion,Mackereth_2017}.
At the extreme of this relation is the ``thick disc'', a population of low-metallicity, highly-alpha-enhanced stars with a vertical scale height of over 1 kpc \citep{Gilmore_1983} (compared to $\sim300$\,pc for the ``thin disc'' to which the Sun belongs), though its status as a chemodynamically distinct component of the Milky Way is debated \citep{Bovy_2012NoThickDisk}.
Thick disc stars do host planetary systems, albeit at a reduced rate \citep{Hallatt_2025}, making the detection of ISOs from this population an interesting possibility, as their composition may on average vary from objects formed around thin disc stars \citep{Lintott2022}. However, the kinematics of the currently-known ISOs are consistent with a thin disc origin for all three, though this has been debated for 3I/ATLAS  \citep{Kakharov_2024,Guo2025,Hopkins2025ATLAS,PerezCouto2025}. Note that classification of stars into the thin or thick disc based purely on kinematics results in significant contamination \citep{Alinder2025}, complicating any relation between an ISO's composition and its kinematics, and estimates of the age of 3I/ATLAS have uncertainties of many Gyr \citep{Hopkins2025ATLAS,Taylor2025}.}

\red{Stellar metallicities and abundances also depend on their formation location, as well as age \citep{Chiappini2001,Nordstrom_2004}. Therefore, the chemical composition of  interstellar objects and the  protoplanetary discs in which they form \citep{Cabral_2023} should vary with both age and formation location within the Galaxy.
In principle, ISOs with different Galactic velocities should  trace the variable compositions of planetesimals formed as a function of  stellar population, which could be probed statistically with Rubin/LSST.}

\red{Using ISOs to probe planetesimal formation does have several issues. One }is that their size, shape and volatile content may have changed during the ejection process. 
This is because ISOs could have been tidally disrupted during ejection, for example if they passed close to a giant planet or star and were ejected \citep{Cuk2018, Rafikov2018, Zhang2020Oumuamua, Veras2020ISOs}. \red{Other issues include thermal processing in their birth system before or during ejection \citep{Gkotsinas2024}, and exposure to cosmic ray radiation during their sojourn through the Galaxy \citep[e.g.,][]{Jackson:2021}.}

\vspace{12pt}

\section{Stellar winds}
\label{sec:winds}

\red{Cometary nuclei and tails are affected by stellar winds. 
They are produced when material from the outermost layer of a star is subject to significant acceleration, overcoming the hold of the star's gravity. Stellar winds are frequently characterised by their acceleration mechanism~\citep{1999isw..book.....L,2023IAUS..370.....V,2023IAUS..370....3O,Vink2026}. }

\red{Massive stars have radiation-driven winds, where the star's luminosity is so great that the radiation pressure blows material outwards. This includes O0--B5 stars, with the winds of B3--B5 stars being somewhat weaker. 
Stars with stellar types B6--A9 do not exhibit significant winds (this includes the exocomet host \(\beta\)~Pictoris, which is an A6V star). 
FGKM-dwarfs exhibit coronal winds, powered by the high temperatures (in the megakelvin range) in the coronae of these stars. 
The stellar types F0--F4 exhibit weak coronal winds due to their narrow convection zones. This category of winds includes the solar wind, as the Sun is a G2V star. Finally, stars that have aged beyond the main sequence have winds driven by dust and pulsations. This information is summarised in Table~\ref{tab:stellar-winds-by-type}.}
\begin{table}

    \newcommand{\TypeRange}[2]{%
     \makebox[1.5em][l]{#1}\;--\;\makebox[1.5em][r]{#2}%
    }

    \newcommand{\MassRange}[4]{%
     \makebox[0.7em][c]{#1}\makebox[1.7em][c]{#2}\makebox[0.7em][c]{#3}\makebox[1.7em][c]{#4}%
    }

    \newcommand{\TempRange}[4]{%
     \makebox[0.7em][c]{#1}\makebox[2.6em][r]{#2}\makebox[1.2em][c]{#3}\makebox[2.5em][r]{#4}%
    }
    \centering
    \caption{
        Summary of broad stellar wind characteristics across spectral types and 
        evolutionary stages. 
        The `Type' column gives the spectral type and subtype range. 
        The `Stage' column describes the evolutionary phase of the star. 
        The `Mass' and `Temperature' columns list the ranges of initial masses and effective temperatures for stars in each category. 
        The `Wind type' column describes the type of wind produced.}
    \begin{tabular}{@{}llccl@{}}
    \toprule
    \red{Spectral}          & Stage             & Mass                                  & Temperature                               & Wind type             \\
    Type              &                   & ($M_\odot$)                           & (K)                          &                       \\ \midrule
    \TypeRange{O0}{B2}        & Main Sequence     & \MassRange{}{}{$\gtrsim$}{$8.0$}      & \TempRange{}{}{$\gtrsim$}{$20000$}        & Hot star              \\
    \TypeRange{B3}{B5}        & Main Sequence     & \MassRange{$\sim$}{$8.0$}{--}{$4.0$}  & \TempRange{$\sim$}{$20000$}{--}{$15000$}  & Hot star (weak)       \\
    \TypeRange{B6}{B9}        & Main Sequence     & \MassRange{$\sim$}{$4.0$}{--}{$3.0$}  & \TempRange{$\sim$}{$15000$}{--}{$10000$}  & Negligible            \\
    \TypeRange{A0}{A9}        & Main Sequence     & \MassRange{$\sim$}{$3.0$}{--}{$1.5$}  & \TempRange{$\sim$}{$10000$}{--}{$7500$}   & Negligible            \\
    \TypeRange{F0}{F4}        & Main Sequence     & \MassRange{$\sim$}{$1.5$}{--}{$1.1$}  & \TempRange{$\sim$}{$7500$}{--}{$6500$}    & Chromospheric (weak)         \\
    \TypeRange{F5}{M9}        & Main Sequence     & \MassRange{$\lesssim$}{$1.1$}{}{}     & \TempRange{$\lesssim$}{$7150$}{}{}        & Chromospheric         \\\midrule
    \TypeRange{K0}{M5}        & Evolved           & \MassRange{}{}{$\gtrsim$}{$8.0$}      & \TempRange{$\sim$}{$5000$}{--}{$3000$}    & Giant star            \\
    \TypeRange{K0}{M5}        & Evolved           & \MassRange{$\lesssim$}{$8.0$}{}{}     & \TempRange{$\sim$}{$5000$}{--}{$3000$}    & Chromospheric         \\
    \TypeRange{M6}{M9}        & Evolved           & \MassRange{$\lesssim$}{$8.0$}{}{}     & \TempRange{$\lesssim$}{$3000$}{}{}        & Giant star            \\\midrule
    $\beta$ Pictoris (A6) & Main Sequence & $1.8$                           & ${\sim\,}8000$                                    & Negligible            \\ 
    Sun (G2)      & Main Sequence     & $1.0$                                 & $\phantom{\sim\,}5778$                                    & Chromospheric         \\\bottomrule
    \end{tabular}
    \label{tab:stellar-winds-by-type}
\end{table}

\red{The composition of stellar winds generally mirrors the composition of the outermost stellar layer but is modified by factors including ionization processes,
mixing and enrichment from the star’s interior,
elemental fractionation \citep[e.g., first ionisation potential effect,][]{Laming_2015}, and
dust formation (in cool stars).}


\red{Winds can be described in terms of density \(\rho\), velocity \(v\), magnetic field \(B\) and temperature \(T\)~\citep[e.g.\ ][]{2021LRSP...18....3V}.
The mass flux of the wind through a closed surface $S$ containing the star  \(\dot M\) is a function of the wind density \(\rho\) and velocity \(v\),}
\begin{equation}
\dot M = \oint_S \rho v \cdot dS.
\end{equation}
\red{In a situation of spherical symmetry the flux simplifies to \(\dot M(r) = 4\pi r^2 \rho v\) where \(r\) is the distance to the stellar centre. For a steady wind there can be no accumulation of matter in the wind so the flux must be independent of \(r\). This imposes a constraint on the density and velocity of the wind}
\begin{equation}
\dot M = 4\pi r^2 \rho(r) v(r)= \text{constant in }r.
\end{equation}
\red{In general the winds accelerate away from the stellar surface and approach a terminal velocity \(v_\infty\) at large distances from the star. Different wind acceleration mechanisms are possible and yield different velocity and density profiles \(v(r)\) and \(\rho(r)\). 
As it expands, the wind emanating from a star generally achieves supersonic speeds.
The velocity profile \(v(r)\) of evolved stars and high-mass stars is often described by a ``\(\beta\) law'' where \(v(r)\propto (1-R_\star/r)^{\beta}\) where \(R_\star\) is the stellar radius (see Sections~\ref{sec:early-type-stars} and \ref{sec:evolved-type-stars}). The velocity profile of the solar wind is described by the~\cite{Parker1958} wind model~(see Section~\ref{sec:solar-type-stars}) and a host of extensions and modifications. For the~\cite{Parker1958} wind model the wind speed at large distances grows as \(\ln r - \ln (\ln r)\), and thus never reaches a terminal velocity.}

\red{Typical ranges of mass loss rates, wind speeds, and pressures around different stellar types are tabulated in Table~\ref{table:stellar_winds}.} \red{Additional information on the dependence of stellar winds on host star mass, including a compilation of empirical formulae for the mass loss rates, can be found in the Appendix. On the Main Sequence the main influence of the stellar wind is on the formation of the cometary tail from outgassed material, with some influence on the surface as discussed in Section~\ref{sec:evolution}. Once stars leave the Main Sequence, the strong winds significantly change the mass of the star itself, with important implications for all orbiting bodies, as discussed next.} 

\begin{table}
    \sisetup{range-phrase = --}%
    \centering
    \caption{ %
    Mass loss rates and terminal wind speeds for different types of stellar winds. The `Pressure at 1 au' column gives the wind ram pressure \(\frac12 \rho v^2\), assuming a steady, uniform wind and that \(v_\infty\) has been reached at a distance of 1 au. In this case $\dot M=4\pi r^2 \rho v$ and the pressure becomes $p = \frac12 \frac{\dot M v}{4\pi r^2}$.
    The following references are used:
    $^1$\citet{Kudritzki2000};  
    $^2$\citet{Cranmer2019};  
    $^3$\citet{2024ApJ...967..120W};  
    $^4$\citet{2022IAUS..366..165H};  
    $^5$\citet{2021LRSP...18....3V};  
    $^6$\citet{Bruhweiler1991}.  
    }
    \label{table:stellar_winds}
    \begin{tabular}{lccc}
    \toprule
    Wind type & Mass loss rate \(\dot M\) & Speed \(v_\infty\) & Pressure at 1 au \\ 
         & (M\(_\odot\)/yr) & (km/s) & (Pa) \\ 
    \midrule
    Hot star (O0--B5)$^1$   & \numrange{e-7}{e-6}    &    \numrange{1000}{3000} & \numrange{e-2}{e-1}\\ 
    Coronal (FGKM)$^2$    & \numrange{e-16}{e-12}  &     \numrange{100}{1500} & \numrange{e-12}{e-7}\\ 
    Giant star (RGB)$^3$ & \numrange{e-11}{e-9} & \numrange{20}{40} & \numrange{e-8}{e-6}\\  
    Giant star (AGB)$^4$ & \numrange{e-7}{e-5}    &     \numrange{5}{30} & \numrange{e-5}{e-2}  \\  
    \midrule
    Solar$^5$ & \(2 \times 10^{-14}\) & 300-800 & \numrange{e-10}{e-9}\\ 
    \(\beta\) Pictoris (A6V)$^6$ & $1\times10^{-14}$ & 60 & $10^{-10}$ \\
    \bottomrule
    \end{tabular}
\end{table}


\section{Evolution of small bodies during the star's giant branch evolution}
\label{sec:GB}

Once a star leaves the Main Sequence, it undergoes significant changes: large increases \red{in} luminosity and radius on the red giant branch (RGB) and asymptotic giant branch (AGB), strong mass loss especially at the AGB tip, and then a declining luminosity as a WD. These changes to the star have significant effects on orbiting bodies: both their physical evolution and their orbital evolution, and thence the global orbital dynamics of the entire system. We discuss these in the closing sections of this \red{work}.

Stars undergoing the giant branch phases of evolution will lose at least half of their mass, expand in size to the orbit of \red{our Solar System's} Earth or Mars, and increase their luminosity by many orders of magnitude. This means that physical effects on orbiting bodies during the giant branch phases are ``supercharged'' compared to main-sequence systems because of the significant changes to the stellar mass, radius and luminosity. The most extreme mass loss, radius and luminosity values are attained during the second giant phase (the AGB), rather than the first (the RGB).

Considering collisions between a body and the star, the increased size of the giant branch star drastically changes the minimum survivable orbital \red{periapsis} of a small body from about a solar radius to about 1-5 au. The envelope boundary of the star is likely to be a good representation of the maximum destruction distance. Small bodies are sufficiently small that tidal orbital decay would not increase this distance, whereas for giant planets the minimum radius for survival can be significantly beyond the maximum stellar radius  \citep{Mustill2012,Adams2013,Madappatt2016}. However, for small bodies radiation effects will likely result in destruction beyond the stellar radius, as discussed below.

RGB and especially AGB stars have strong winds (see Appendix~\ref{sec:evolved-type-stars}), owing to an interplay of low surface gravity, stellar pulsations, dust formation, and high radiation pressure \citep{2022IAUS..366..165H}. Mass loss from the host star will have both direct physical effects on the bodies, and indirect gravitational effects on their orbits. 
The gravitational effect arises as the orbit evolves in response to the star's changing gravitational potential. 
If the body is close enough to the star (within a few hundred au of the parent star, so that the orbital timescale is short compared to the mass-loss timescale\footnote{\red{This is s}ometimes known as ``adiabatic'' mass loss.}), then the orbit will expand by a factor of a few, while maintaining its shape in the form of eccentricity, inclination and  orientation in space \citep{Hadjidemetriou1963,Omarov1964}. 
However, for exo-Oort cloud bodies on wider orbits, the orbit changes shape and may shrink\footnote{As measured by the semimajor axis. For isotropic mass loss, the orbital \red{periapsis} distance always increases \citep{Veras2011}.} or expand depending on both its initial shape and the location of the body along the orbit \citep{Veras2011,Veras2012}. 
Significantly anistropic stellar mass loss can give rise to more complex changes in the orbit \citep{Parriott1998,Veras2013,Stone2015,Dosopoulou2016First,Dosopoulou2016Second,Akiba2024}. 

The stellar mass loss takes place through a strong wind (Appendix~\ref{sec:evolved-type-stars}), but the physical and chemical consequences of giant branch stellar winds impinging on the surface of a body have not yet been significantly explored. \red{\cite{Veras2015} presented a general formalism for the orbital effects of drag from giant branch stellar winds, finding it to be significant, though detailed treatments including inhomogeneities in the wind have not been performed}. Radiative effects are, however, much better studied\red{; see \cite{Vrignaud2026} for a discussion}. Stellar luminosities on the giant branch attain values of thousands of solar luminosities, and this has extreme effects on heating, sublimation and devolatilisation of bodies \citep{Jura2010,Jura2012,Malamud2016,Malamud2017a,Malamud2017b,Katz2018,Harrison2021,Li2024}. The result of these processes could fundamentally change the shape, chemistry and survival prospects for these bodies. Furthermore, such extreme radiation will change the spin and orbit evolution of these bodies through a supercharged version of the YORP \citep{Veras2014YORP,Veras2020YORP} and Yarkovsky \citep{Veras2015,Veras2019} effects (see Section~\ref{sec: nonGravAndTidalForces}), leading to both widespread destruction due to rapid and irreversible rotational spin-up, and orbital \red{dispersal} of the survivors and fragments. \red{The Yarkovsky \red{effect} dominates over the Poynting--Robertson (PR) effect, an additional radiative force \citep{Burns1979,Gustafson1994}, for bodies orbiting giant branch stars \citep{Veras2015}, though the PR effect may help bring small (sub-m) fragments down to the surface of a young white dwarf \citep{Veras2022}.} 

\red{Direct evidence for these processes occurring is lacking, although t}he central star of the Helix Nebula shows a strong IR excess that may be due to a large population of disintegrating \red{small} bodies \citep{Marshall2023}, while the central star of the planetary nebula WeSb~1 has recently been observed to undergo occultations attributed to dust clouds \citep{Budaj2025}. \red{On the other hand, t}he white dwarfs that emerge after the AGB phase show significant evidence for cometary phenomena arising from some form of small bodies---whether volatile-rich or volatile-poor---which likely stems ultimately from the aforementioned changes to their orbits. This is discussed in depth in the next Section.


\section{Cometary activity and related phenomena close to white dwarfs}
\label{sec:WD}

\begin{figure}
    \centering
    \includegraphics[width=0.9\textwidth]{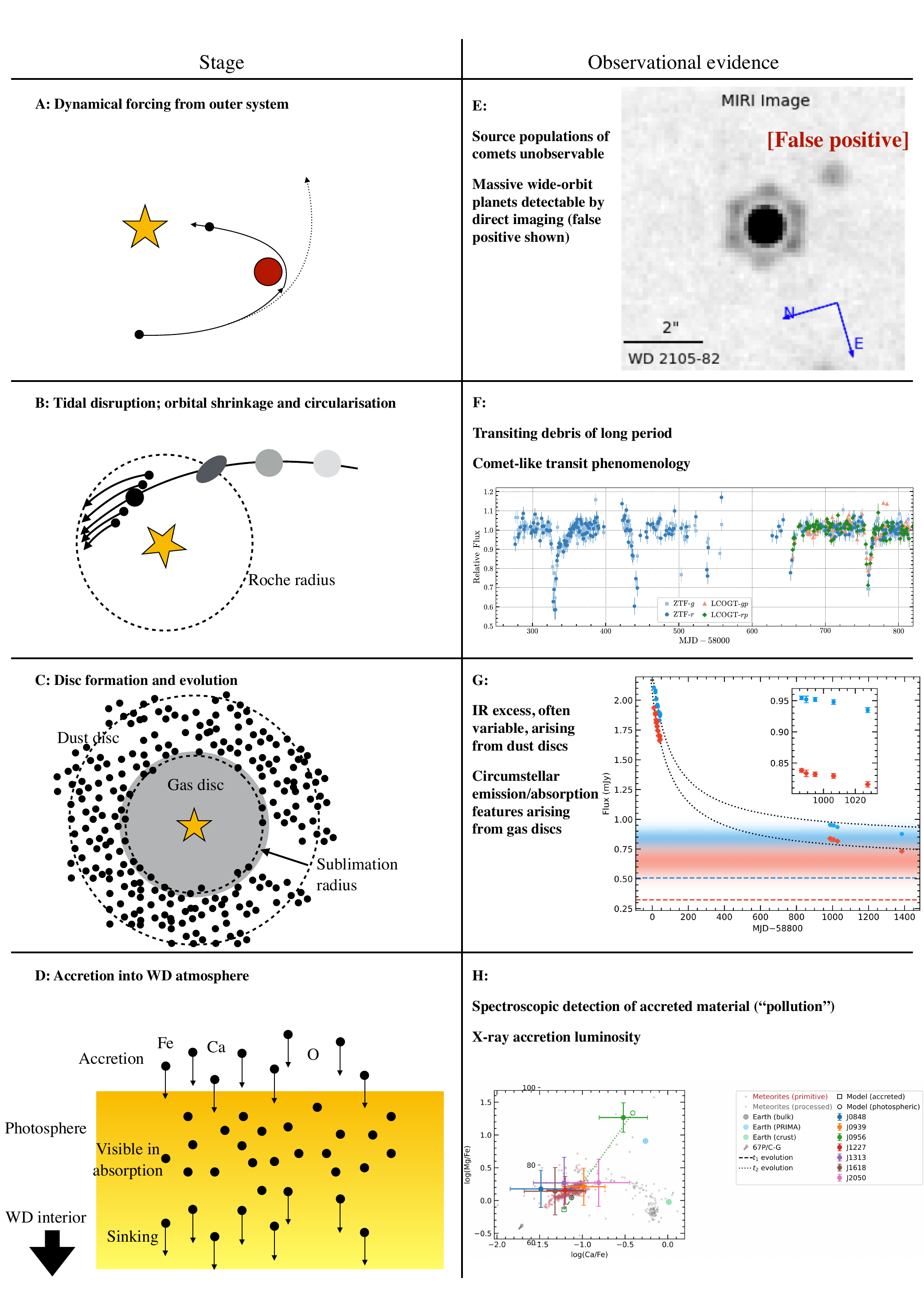}
    \caption{Stages in the delivery of small bodies to WDs (A--D), along with observational evidence for each stage (E--H). 
    Observations shown are: (E) \red{False-positive} Super-Jovian planetary candidate companion to 
    WD~2105-82 imaged at 15 microns with JWST/MIRI \citep[][Fig.\ 1$^*$]{Mullally2024};
    (F) Deep, asymmetric occultations by material on long-period, possibly eccentric orbit around ZTF~J0139+52 \citep[\red{Figure courtesy Zach Vanderbosch, based on}][Fig.~1]{Vanderbosch2020}; (G) Decrease \red{with time} in the 3- and 5-micron IR flux excess (Spitzer/IRAC and JWST/NIRSpec) of WD~0145+234, possibly due to collisional evolution of the dust disc \citep[][Fig.\ 1$^*$]{Swan2024}; (H) Spectroscopically-inferred abundances of material accreted onto 7 WDs, compared to Solar System bodies \citep[][Fig.\ 1$^*$]{Swan2023}. \\ $^*$Reproduced under the CC-BY-4.0 license \url{https://creativecommons.org/licenses/by/4.0/}}
    \label{fig:WD-close}
\end{figure}

As a result of the changes discussed in Section~\ref{sec:GB}, a young white dwarf (WD) is expected to have no planets or small bodies within at least a few au, owing to the combination of stellar mass loss, tidal orbital decay, and strong radiation effects. However, a high fraction of WDs do in fact possess some material in very close proximity to the star, within roughly $1\mathrm{\,R_\odot}$. The evidence for this comes in several forms: photometric transits of material and/or parent bodies across the stellar disc; dusty circumstellar material inferred from an IR excess; circumstellar gas discs inferred from emission or absorption lines; and \red{metals} in the stellar photosphere deposited by infalling material, and seen in absorption (Figure~\ref{fig:WD-close}). These phenomena are thought to arise as a result of bodies on wide orbits being destabilised around the WD following stellar mass loss. Unfortunately, the source regions are difficult to probe observationally (Fig~\ref{fig:WD-close}A,E)\red{. Some giant planet candidates have been imaged with JWST \citep{Mullally2024}, but subsequent epochs have revealed that they were false positives \citep{Mullally2026}}\footnote{\red{An earlier discovery, a super-Jovian companion to WD0806-661 at 2500\,au projected separation, is often regarded as a brown dwarf owing to its large mass and very large separation \cite{Luhman2011,Rodriguez2011}.}}. WDs typically descend from progenitors somewhat more massive than the Sun, as most solar-mass stars have not yet had time to evolve off the Main Sequence \citep{Koester2014}. Their progenitors are therefore often A-type stars, which are known to host debris discs very commonly \red{\citep[around a quarter or a third show detectable IR excess,][]{Su2006,Thureau2014,Bannister2025}}, but little is known about the planets at several to 10s of au that can drive the dynamics. This has given theorists great freedom in speculating on the exact dynamical processes driving the delivery.

For further reading on this topic, see the recent reviews by \cite{Veras2024,Xu2024,Malamud2026,Mustill2026}.

\subsection{Transiting material}

A handful of WDs are now known to possess material that passes in front of the star, detected as transits in broadband photometric lightcurves. Transits of WDs by planets can easily achieve 100\% transit depth owing to the small stellar radius, and transits of small planets, asteroids and comets can be detected much more easily than towards MS stars \citep{Agol2011}. However, the small size of the WD makes the a priori geometric probability of a transit small. An early search for transits of WDs in WASP data was unsuccessful \citep{Faedi2011}. The first successful detection was with the \emph{Kepler} space telescope by \cite{Vanderburg2015}, who detected asymmetric transits of WD1145+017. These transits, with their steep ingress and shallower egress, were attributed to a comet-like morphology arising from a parent body possibly surrounded by a coma and with a trailing tail. Several orbital periods were identified at $\sim4.5$\,hr, suggesting multiple bodies close to the Roche radius, possibly indicative of a recent break-up event, and that the cometary tails may arise from tidal disruption, rather than sublimation \citep{Veras2017}. Multi-wavelength photometry of the transits has shown that they are independent of wavelength, posing challenges for our understanding of the dust being released by the parent body: at such small orbital radii, velocities are high and therefore small dust should abound from fragmentation, yet small grains would induce a significant colour dependence to the transits. Possible resolutions are that small grains sublimate extremely \red{quickly} \citep{Xu2018}, or that the dust is sufficiently abundant to be optically thick at all wavelengths \citep{Izquierdo2018}. Recently this system has become more quiescent \citep{Aungwerojwit2024}, showing that these transiting phenomena are transitory on human-relevant timescales. This raises the prospect that revisiting previously-surveyed WDs with non-detections could in future yield more detections, as fresh material is delivered to the WD or existing material which was previously inactive and unseen undergoes more outbursts, collisions, and dust production.

\red{Since the WD1145 transit detections were first recorded}, transits have been detected around several more WDs. Just under $1\%$ of metal-polluted WDs exhibit transits, which is consistent with \emph{all} such WDs having detectable material in orbit once the geometric transit probability is accounted for \citep{Robert2024}. WD1856+534 \citep{Vanderburg2020} is a typical hot Jupiter on a 1.4-day orbit, save for the fact that it orbits a WD rather than a Main Sequence star. \red{ZTF J1944+4557 resembles WD1145 in its period (4.9\,hr), and also became quiescent after discovery, although its transits have subsequently re-appeared \citep{Guidry2025}.} ZTF J0328-1219 \citep{Vanderbosch2021} exhibits recurrent features similar to WD1145, but at longer periods of 9.9 and 11.2\,hr, pointing to sublimation, rather than Roche lobe overflow, as the mechanism for generating the activity\red{; SBSS 1232+563 shows a signal with a comparable 14.8\,hr period, along with a 40\% flux drop lasting for 8 months \citep{Hermes2025}}. ZTF J0139+5245 on the other hand exhibits recurring transits with a long period of 107 days, with a pronounced cometary morphology \citep{Vanderbosch2020}; see Fig~\ref{fig:WD-close}B,F. These likely arise from a body on an eccentric orbit in the process of orbital circulation, be that due to tidal or radiation forces \citep{Veras2015Shrink}; note that, with a period of 107 days, the \red{apapsis} of such an orbit must still be considerably interior to the expected source region at $\gtrsim10$\,au, indicating some orbital decay since the body was injected from the source region. Apart from the planet WD1856b, these transits are all variable with time, in terms of depth and precise morphology, further evidence of the comet-like, rather than solid, nature of the occluding material.

Perhaps the most complex transiting system yet found is WD1054-226 \citep{Farihi2022}. In this system, there is no flat, out-of transit baseline; rather, the stellar light curve is constantly varying, with two dominant periods: one at 25 hours, and one at 23 minutes. These lie on a 65:1 commensurability, suggesting some kind of resonant interaction between a disc of material and a perturbing planet or asteroid\footnote{\red{The 65:1 commensurability is between the observed periods; this likely implies not a 65:1 mean motion resonance, but a 66:65.}}. This could involve the disc material being concentrated in resonant clumps \citep[e.g.,][]{Wyatt2003}, or the perturber exciting a standing wave in the disc as seen in ring--moon interactions in the Saturnian system \citep[e.g.,][]{Cuzzi1981}, although models here are lacking. On short time-scales (night-to-night), the detailed light-curve features such as double-dips align well when phase-folded on the \red{25}-hour period; on longer time-scales \red{of several years}, the detailed features vary but the dominant frequencies remain \red{\citep{Farihi2022,Korth2026WD}}. 

\red{The discovery of these systems is now accelerating, with \cite{Bhattacharjee2025} presenting 6 new candidates from an automated analysis of Zwicky Transient Facility (ZTF) light curves. One of these systems, WD J1013-0427, shows the first reddened transits of dust clouds in front of a white dwarf, providing evidence for sub-micron grains\footnote{\red{Previously, \citep{Hallakoun2017} had detected bluing during transits of WD1145+017, attributed to gas absorption during transit rather than dust properties.}}. This is in contrast not only to WD1145+017 \citep{Alonso+2016,Izquierdo2018,Xu2018} but also to several other WDs which also exhibit grey transits: WD J0328-1219 \citep{Gary2024}, ZTF J1944+4557 \citep{Guidry2025}, SBSS 1232+563 \citep{Hermes2025}, and WD1054-226 \citep{Farihi2022}; especially in this latter case, the lower temperature of the dust means that the lack of observed small grains cannot be attributed to their sublimation \citep{Korth2026WD}.}

\subsection{Circumstellar discs}

Discs orbiting WDs can comprise dust, gas, or both. The observed discs all lie within $\sim1\mathrm{\,R_\odot}$, consistent with the outcome of tidal disruption and subsequent orbital decay. The descendants of \red{main-sequence} debris discs, on much wider orbits, are unfortunately mostly undetectable owing to the extremely low luminosities of WDs \citep{Bonsor2010}. An exception may be the debris disc around the very young WD at the centre of the Helix Nebula \citep{Su2007}; this object may indeed be showing cometary activity and the dust could be produced by outgassing, rather than collisions as in typical debris discs \citep{Marshall2023}.

The dust discs orbiting WDs are inferred indirectly from excess IR emission compared to the stellar spectral energy distribution: they cannot be resolved, unlike many MS debris discs, owing to their extremely small radii. \red{Over} 1\% of WDs exhibit such IR excess \red{\citep{Wilson2019,MadurgaFavieres2024,Murillo-Ojeda2026}}. Where good-quality mid-IR spectra exist, mineralogy is possible: analysis of \emph{Spitzer} IRS data showed that the G29-38 disc contains crystalline pyroxene, amorphous silicates, water ice, metal sulphides, and possible amorphous carbon \citep{Reach2009}. The improved sensitivity afforded by JWST is now enabling analyses of many more discs \citep{Swan2024,Farihi2025}. Many IR excesses are variable on timescales of months to years \citep[e.g.,][]{Xu2018Variability,Swan2020,Swan2024}, indicating significant changes in the quantity of dust in the systems (Fig~\ref{fig:WD-close}C,G). The near-WD environment is clearly highly dynamic, with ongoing production of dust grains through fresh delivery and disruption of new parent bodies, collisions, and/or evaporative outbursts.

Discs of circumstellar gas are detected spectroscopically, both in absorption (which requires an unusually favourable geometrical alignment to have the disc edge-on) or emission (which does not). There are fewer detections of gas discs than dust discs, but they do permit detailed measurements of the elemental abundances of the material \citep{Steele2021}. Unlike typical astrophysical gas discs, these are formed from vapourised rock, and are not dominated by H/He. Disc kinematics can be inferred from the Keplerian line profiles, and they sometimes show an orbital eccentricity and relativistic precession \citep{Manser2016}. The disc emission of SDSS J1228 shows variability on a 2-hour timescale, attributed to a disturbance in the disc by an embedded asteroid or minor planet \citep{Manser2019}, orbiting interior to the fluid Roche limit but held together by its internal strength.

\subsection{Accreted material}

\label{sec:WDaccretion}

In retrospect, we can identify the first detection of extrasolar planetary material by Humanity \citep{Farihi2008}: \cite{vanMaanen1917} identified a peculiarly faint high-proper motion star of spectral class F0, which we now know to be the first discovery of a metal-polluted white dwarf.

Metals are not expected to be present in most WD atmospheres, since the strong surface gravity should cause material heavier than H/He to sink below the photosphere on astrophysically short timescales (ranging, depending on the atmospheric temperature and whether H or He dominates, from days to $\sim1$\,Myr; see for example \citealt{Wyatt2014}). This implies ongoing or recent accretion of non-stellar material, now thought to arise from the accretion of planets, moons, or small bodies onto the WD (Fig~\ref{fig:WD-close}D). This is an extremely common phenomenon: depending on the sub-population of WDs studied, up to 50\% exhibit spectroscopically detected metal lines \red{\citep[e.g.,][]{Koester2014,Wilson2019,OuldRouis2024}}.

From the elemental abundances measured in WD atmospheres, we can determine the bulk compositions of the progenitor bodies, and compare them to bodies of our own Solar System (Fig~\ref{fig:WD-close}H): readers are referred to the reviews by \cite{Jura2014} and \cite{Xu2024} for the geochemical conclusions. Information obtained in this manner is complementary to our study of the atmospheres of exoplanets via transmission or emission spectroscopy (which tells us about the surface but needs to be coupled to interior models to learn about the interior), and to knowledge of the bulk density from a planet's mass and radius (which provides weak constraints on the bulk composition with many degeneracies). Disadvantages are that this can only be performed for destroyed bodies after the star has become a WD, that we may not know the primordial nature of the parent body (volatile-rich, volatile-poor, etc.), that the material has undergone thermal processing on the RGB/AGB altering its composition to some degree (see Sec~\ref{sec:GB}), and that not all components of the destroyed parent body may be accreted at the same time and rate \citep{Brouwers2023}.

Currently over 1700 polluted WDs have been identified \citep{Williams2024}. This number is expected to grow rapidly in the near future, as the combination of large-scale spectroscopic surveys such as SDSS, J-PAS, LAMOST and \emph{Gaia,} combined with machine learning techniques, allows for efficient screening of vast numbers of spectra to identify candidate polluted WDs and efficiently target high-resolution follow-up spectroscopy for confirmation and to increase the number of species detected \citep{Kao2024,Perez-Couto_2024}. Future surveys such as 4MOST \citep{deJong2019} will soon provide over $10^5$ high-quality WD spectra for large-scale studies of WD abundances \citep{Chiappini2019}.


While we typically infer the accretion of material from planets or small bodies indirectly via the presence of metals in the WD photosphere, direct evidence of accretion has been found at G29-38 in the form of X-rays emitted by cooling of the accreted material \citep{Cunningham2022}.

\subsection{Dynamical evolution around WDs}

\begin{figure}
    \centering
    \includegraphics[width=0.99\textwidth]{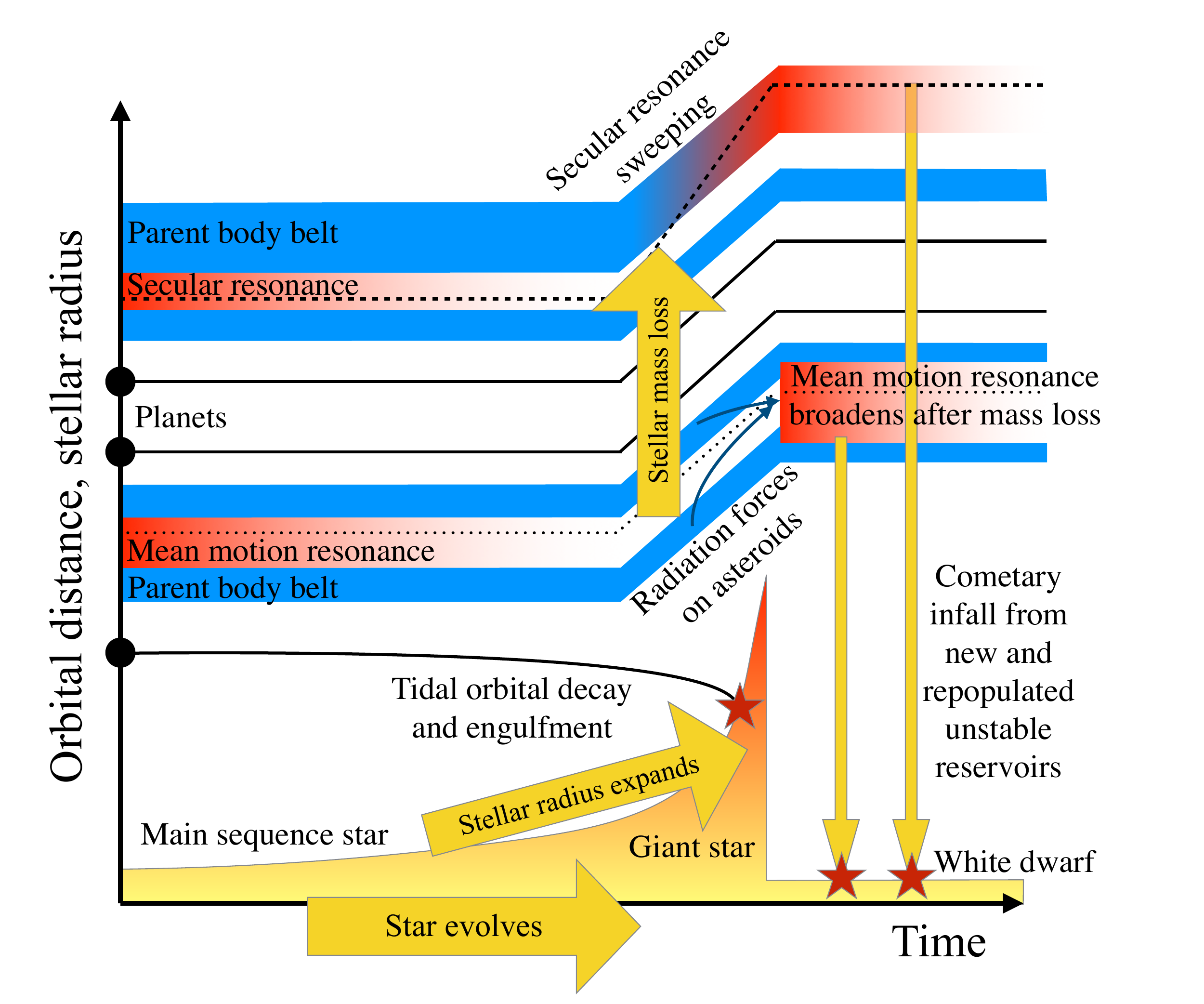}
    \caption{Schematic overview of how changes wrought by stellar 
    evolution effect the delivery of small bodies to
    white dwarfs. We show an imaginary system initially of three planets 
    with two source belts, one destabilised by a mean-motion resonance, 
    and one destabilised by a secular resonance. As a result of stellar 
    mass loss, the MMR broadens, while previously stable bodies 
    are moved by radiation forces into the unstable region. The secular resonance
    moves as a result both of the engulfment of one planet by the giant star, and of 
    the differential motion of the remaining planets as tidal forces compete
    with the effects of mass loss. }
    \label{fig:WD-stability}
\end{figure}

The above phenomena are thought to reflect stages in a process of dynamical excitation of the eccentricities of the parent bodies. The initial excitation may arise from the mechanisms discussed in Section~\ref{sec:depletion}, such as scattering, forcing from mean motion or secular resonances, or the Lidov--Kozai effect. With sufficiently extreme eccentricity excitation, a body's \red{periapsis} is forced down to very small values, after which it may experience tidal disruption, circularisation, pulverisation, sublimation, and ultimately accretion onto the WD. A schematic of these processes is shown in Figure~\ref{fig:WD-close}. 

Delivery of bodies to the region close enough to the WD for the above phenomena to occur is possible for bodies originating on wide \red{orbits similar to the Solar System's scattered disc or Oort Cloud objects \citep{Alcock1986} These} orbits are affected by near-impulsive anisotropic mass loss as the star transitions to a WD, known as WD ``kicks'', and then continue to evolve under the Galactic tide and stellar flybys \citep{Veras2014Galactic,Stone2015,Veras2014Comet,OConnor2023,Pham2024} and in some cases internal interactions within the planetesimal population \citep{Akiba2024}. While \red{(exo-)Oort Cloud} dynamics \red{are} identical for \red{bodies} orbiting a MS star or a WD, the combination of reduced stellar mass and increased orbital radius means that a body \red{experiences} a stronger effect from the Galactic tide following AGB mass loss. The presence of a planet or star between the Oort Cloud and the WD can reduce delivery of Oort Cloud material to the WD, as they can suppress eccentricity fluctuations induced by the Galactic tide and eject bodies that do get too close, preventing them from crossing the planet's orbit \citep{OConnor2023,Pham2024}. Overall, exo-Oort cloud bodies may be responsible for the trace H seen in many He-atmosphere WDs, but are unlikely to be the dominant source of metal pollution \citep{Veras2014Comet}.

Initial reservoirs of the parent bodies of comets are slowly depleted through a star's MS lifetime by dynamical and collisional processes (Sec~\ref{sec: mechanismsToGetCometsCloseToStars}, \ref{sec:depletion}). The strong mass loss at the end of a star's life can destabilise formerly stable systems or reservoirs, and radiation forces can repopulate depleted reservoirs by moving bodies in the system (Section~\ref{sec:GB} and Figure~\ref{fig:WD-stability}). As a star loses mass, planet:star mass ratios increase, making interactions between planets and minor bodies stronger than previously when compared to the dominant interactions between planets and the host star \citep{Duncan1998,Debes2002}. The stability of a multi-planet system is governed by the planet:star mass ratio $(M_\mathrm{pl}/M_\star)^\alpha$, where $\alpha=1/3$ for a two-planet system \citep{Gladman1993} and $\alpha=1/4$ for a three-planet system \citep{Quillen2011,Petit2020}. Entire systems which survived the MS in stable configurations can therefore undergo scattering after stellar mass loss, in the process scattering small bodies down to the WD \citep{Mustill2018}. In more quiescent cases, an unstable ``Kirkwood gap'' around a mean-motion resonance can increase in width \citep{Debes2012}, \red{as can} the chaotic region of overlapping resonances \red{close to a planet} \citep{Bonsor2011,Caiazzo2017}. In systems where one or more planets undergo tidal decay and/or engulfment, unstable secular resonances will move around the system, destabilising fresh, previously stable, reservoirs \citep{Smallwood2021}. Radiation forces on small bodies can also move the bodies into unstable regions (See Sec~\ref{sec:GB}). These processes are reviewed in detail \red{by} \cite{Mustill2026}.

Unfortunately, the large range of parameter space for setting up multiplanet systems, coupled with the lack of data on planets on wide orbits, means that there are many possible mechanisms with little observational guidance to choose between them. We do know that WD pollution is common, and that the accretion rates are relatively weakly decaying, over Gyr timescales \citep{Blouin2022}. This rules out instabilities in systems of giant planets as the dominant driver, as these planets are uncommon and the instabilities occur too quickly to explain late pollution \citep{Mustill2014,Mustill2018}. Forcing from stellar binary companions can in principle explain late pollution \citep{Hamers2016,Petrovich2017}, but observationally WD pollution does not seem to correlate with binarity \citep{Wilson2019,Noor2024}.

Currently, two promising delivery mechanisms involve systems of high planetary multiplicity: destabilisation of high-multiplicity systems by mass loss may be common \citep{Maldonado2021}, and if the major planets remain stable there is scope for secular resonances to be moved around or repopulated during the AGB \citep{Smallwood2021}, so they can act on fresh reservoirs of material.

\section{Open Questions}

\label{sec:Qs}

The following are some of the major open questions in the field of exocomet evolution. Future studies answering these would help us understand these objects and the processes occurring in their planetary systems.
 
\begin{itemize}
    \item What \red{processes drive} comets down to small \red{periapsides} around MS stars? Are the main driving bodies planets, stellar binary companions, or the Galaxy as a whole, and what are the exact dynamical pathway(s)?
    \item What process drives comets down to small \red{periapsides} around WDs? Is this the same as for MS stars, or do different mechanisms dominate? In both of these cases, we are currently hindered by our lack of knowledge of the architectures of outer regions of planetary systems.
    \item Are certain planetary system architectures required to produce low-\red{periapsis} exocomets, or are a range of architectures and mechanisms possible?    
    \item How can we observationally probe planets in the outer regions of planetary systems, where comets probably originate? 
    \item What is the relative importance of stellar flybys and the tidal field for exo-Oort Cloud depletion? How does this change depending on a star's migration history within the Galaxy?
    \item What surprises can we expect from the upcoming census of ISOs from \red{The Vera C.\ Rubin Observatory's Legacy Survey of Space and Time? What will be the balance between more traditional objects such as 2I/Borisov and 3I/ATLAS, and more exotic objects such as 1I/'Oumuamua?}
    \item How unevolved are the nuclei of dynamically new comets? What insights can we expect from Comet Interceptor?
    \item Is the physical evolution of exo-cometary nuclei the same as that for Solar System comets? How can we probe this when the individual bodies themselves are too small to detect?
    \item \red{To what extent do asteroids on cometary orbits, low-activity comets, main-belt comets, and related populations represent distinct reservoirs versus different evolutionary states of a common feedstock?}
    \item What is the primordial nature of the bodies responsible for WD pollution? Were they originally volatile-poor, or have they become so by devolatilisation when the star was a giant?
\end{itemize}

\section{Conclusions}
\label{sec:conclude}

Comets appear to be a common phenomenon not just in our Solar System, but also in other planetary systems, both on the main sequence and around white dwarfs. The active cometary phase is transient, and to bring it about a body must be moved dynamically from its orbit in a source reservoir with a large \red{periapsis} to an orbit with a smaller \red{periapsis}. While these dynamical processes are well understood in the Solar System, with different dynamical channels giving rise to different families of comets, when it comes to most extrasolar systems we must fall back on speculation, as the planets that might be responsible for driving the dynamics are mostly undetectable. A range of dynamical processes may contribute in different measure, and each system is likely to be unique in its details. These processes include Galactic tides, stellar flybys, direct gravitational scattering by planets, eccentricity forcing in mean-motion and secular resonances, and the von Zeipel--Lidov--Kozai effect induced by an inclined planet or binary stellar companion;  multiple processes may be active in any one system, whether on different bodies or even the same body (for example, planetary scattering following resonant eccentricity forcing). Many of these mechanisms can not only deliver bodies to the inner regions of a planetary system to become comets, but also liberate them from the system entirely to become free-floating interstellar objects such as 1I/`Oumuamua\red{, 2I/Borisov, or 3I/ATLAS}.

In contrast to our lack of precise knowledge about the planetary demographics perturbing cometary source reservoirs, we have a reasonably good understanding of the demographics and evolution of massive Kuiper Belt analogues, which are visible in the infrared and submillimetre as debris discs. These are seen around main sequence stars of a range of spectral classes. A large number of debris discs have now been spatially resolved, especially with Herschel in the far-IR and ALMA in the sub-mm, and in many cases their morphology suggests ongoing dynamical interactions with one or more planets in the system, even though these have not yet been directly detected.

For our understanding of the evolution of the cometary nuclei themselves, we are largely reliant on Solar System observations and extrapolation of these models to exocomets. Non-gravitational forces, arising from sublimation and irradiation near \red{perihelion}, have been directly observed to change the spins and orbits of small bodies in the Solar System. In our Solar System, we can observe cometary bodies at all stages of their evolution, from storage in the outer Solar System through to activity \red{as they warm up when approaching the Sun,} followed by eventual break-up, and with a number of comets now visited it is clear that the geology of these bodies can differ greatly, likely at least in part depending on how long they have been on their low-\red{perihelion} orbits. In the near future, it is hoped that the Comet Interceptor mission will be able to opportunistically study up close a dynamically new comet on its first \red{perihelion} passage through the inner Solar System, thus investigating a \red{more} pristine object. \red{Our sample of ISOs is so far small, but has thrown up one major surprise in the unusual, possibly highly N$_2$-rich, body 1I/`Oumuamua.}

Finally, a large number of white dwarfs show comet-like phenomena such as asymmetric transits. Deposition of material into the WD atmosphere provides an important means of probing the bulk composition of these objects, complementing other means of studying the compositions of extra-solar bodies such as planetary transmission spectroscopy. The material deposited into WD atmospheres appears to be largely depleted in volatiles, but it is as yet unclear whether these are on the whole actually volatile-poor asteroids, or \red{formerly} icy bodies more like Trans-Neptunian Objects that have been devolatilised by the extreme luminosity increase that their host star underwent during its giant branch evolution.



\backmatter



\bmhead{Acknowledgements}

We gratefully acknowledge support by the International Space Science Insitute, ISSI, Bern, for supporting  and hosting the workshop on ``Exocomets: Bridging our Understanding of Minor Bodies in Solar and Exoplanetary Systems'', during which this work was initiated in July 2024. \red{We thank two anonymous reviewers for their constructive comments which improved the paper.}

M.T.B. appreciates support by the Rutherford Discovery Fellowships from New Zealand Government funding, administered by the Royal Society Te Ap\={a}rangi. M.J.H. acknowledges support from the Science and Technology Facilities Council through grant ST/W507726/1. K.J. acknowledges support by grants from eSSENCE (grant number eSSENCE@LU 9:3), the Swedish National Research Council (project number 2023-05307), The Crafoord foundation and the Royal Physiographic Society of Lund, through The Fund of the Walter Gyllenberg Foundation. R.K. would like to acknowledge the support from ``L'Oreal UNESCO For Women in Science'' National program for Bulgaria. This work benefited from discussions within ISSI International Team project 504 “The Life Cycle of Comets” funded by the International Space Science Institute (ISSI) in Bern. R.K. acknowledges partial support by project KP-06-D002/3 “CLIC – Cometary Life Cycle,” carried out under the PROMYS (Promotion of Young Scientists) component of the Swiss–Bulgarian Research Programme and supported by the Bulgarian National Science Fund. A.J.M. acknowledges support from the Swedish Research Council (Project Grant 2022-04043) and the Swedish National Space Agency (Career Grant 2023-00146). T.D.P. is supported by a UKRI Stephen Hawking Fellowship and a Warwick Prize Fellowship, the latter made possible by a generous philanthropic donation. X.P.C. acknowledges financial support from the Spanish National Programme for the Promotion of Talent and its Employability grant PRE2022-104959 cofunded by the European Social Fund. \red{D.Z.S. is supported by an NSF Astronomy and Astrophysics Postdoctoral Fellowship under award AST-2303553. This research award is partially funded by a generous gift of Charles Simonyi to the NSF Division of Astronomical Sciences.  The award is made in recognition of significant contributions to Rubin Observatory's Legacy Survey of Space and Time.}

\bmhead{Competing interests}

The authors have no competing interests to declare that are relevant to the content of this article.

\begin{appendices}

\section{Stellar winds -- dependence on host star mass}

\label{app:winds}
\subsection{High-mass (O0--B5) type stars}\label{sec:early-type-stars}
High-mass stars with spectral types O0--B5 have stellar winds driven by the star's radiation pressure. These stars emit copious amounts of ultraviolet light, which exerts pressure on the outermost stellar layer. The radiation pressure primarily acts on metal ions through absorption and scattering in spectral lines, leading to the acceleration of the wind~\citep{Castor1975,2023IAUS..370..130S}.
For these hot stars, the wind velocity profile \(v(r)\) is often~\citep{1999isw..book.....L,2011A&A...534A..97K} described by a \(\beta\) law
\begin{equation}\label{eq:beta-law}
v(r) = v_0 + (v_\infty - v_0) \left(1 - \frac{R_\star}{r}\right)^\beta .
\end{equation}
The \(v_0\) value \red{sets} the wind velocity at an inner boundary but plays \red{little} role at large distances from the star; it is sometimes set to \(v_0=0\)~\citep{2011A&A...534A..97K}. 

The value \(\beta\) determines the acceleration at different stellar distances. \red{A value of} \(\beta=\frac12\) corresponds to an inverse square accelerating force \(F\propto r^{-2}\) \citep{1934MNRAS..94..522C}. Values of \(\beta=\numrange{0.7}{1}\) are adopted for O star winds~\citep{1996A&A...305..171P}, whilst higher values are possible for evolved stars~\citep{2006A&A...446..279C} such as \(\beta\sim2\) for BA-supergiants~\citep{2008A&ARv..16..209P}.
More sophisticated models may vary \(\beta\)~\citep[e.g.\ for Wolf-Rayet stars,][]{2005A&A...432..633G}.

The terminal velocity \(v_\infty\) is often linked to the escape velocity at the stellar surface \red{(adjusted to include the effects of radiation pressure) $v_\mathrm{esc}$}. 
The review by~\citet{Kudritzki2000} uses
\begin{equation}
v_\infty = C(T_\text{eff}) v_\text{esc}, \quad C(T_\text{eff}) = \left\{
\begin{array}{ll}
    2.65 & \;\;\; T_\text{eff} > 21\,000 \text{K}\\
    1.4 & \;\;\; 10\,000 \text{K} < T_\text{eff} < 21\,000 \text{K} \\
    1.0 & \;\;\; T_\text{eff} < 10\,000 \text{K}\\
\end{array}
    \right.
\end{equation}
Hot star winds can reach speeds of several thousand kilometers per second and produce significant mass loss values.

\red{Observationally-determined formulae governing the winds of hot stars can be found in \cite{Vink2001} for O and B stars, and \cite{Nugis2000} for Wolf--Rayet stars.}

\subsection{Weak winds (B6--A9)}\label{sec:weak-winds}

For B6--A9 stars, the hot-star wind mechanisms are not active\red{; nor are} the cool-star wind mechanisms \red{described below}. Theoretical modelling suggests that small convective envelopes can form near the surface of these stars but the winds are expected to be negligible~\citep{2019ApJ...883..106C}.

\subsection{Cool stars (FGKM)}\label{sec:solar-type-stars}
Low-mass stars form stellar coronae, an outer layer of high-temperature matter in the megakelvin range~\citep{2012RSPTA.370.3217P}.
The stellar winds of low-mass stars are driven by the high thermal pressure arising from the high temperature of their coronae. Stellar surface (i.e. photospheric) magnetic fields, and their continuation into the magnetised corona and wind regions, play a crucial role in accelerating and structuring the winds.

Models of cool star winds are based on (magneto-)hydrodynamic models and the conservation of mass and angular momentum, the ideal gas law, and the polytropic \red{equation of state}. 
Under these conditions the one-dimensional velocity profile is independent of the wind mass loss rate \(\dot M\), and given by~\red{\cite{Parker1958,1960ApJ...132..175P}}
\begin{equation}\label{eq:mom_polytropic}
    \frac{\mathrm{d}v}{\mathrm{d}r} = \frac{v}{v^2 - c_\text{s}^2}  \left(
        \frac{2c_\text{s}^2}{r} - \frac{GM}{r^2} 
    \right),
\end{equation}
where \(c_\text{s}\) is the sound speed for a polytropic gas. 
For a more detailed description of cool star winds, see~\citet{2021LRSP...18....3V}.

For an a constant-temperature (isothermal) wind, \(c_\text{s}\) is also constant, and then \eqref{eq:mom_polytropic} is the well-known isothermal 
\citet{Parker1958} 
wind model. In this isothermal case the wind profile can expressed \red{analytically} via the the Lambert~\(\operatorname{W}\) function~\citep[\red{see}][]{2004AmJPh..72.1397C}. 
For \red{the non-isothermal case} there is no closed-form solution to \mbox{Equation \eqref{eq:mom_polytropic}}; instead the equation must be solved numerically. 

Polytropic wind models, such as the Parker model and its extensions, are applied to an `already heated' corona, i.e. these models are separated from the problem of coronal heating. For the Sun and other low mass stars, coronal-heating mechanisms are not well understood.
An extended model that includes coronal heating is described \red{by}~\citet{2018LRSP...15....4G}.

In modelling based on the solar wind, velocities of coronal winds generally range from 300--800\,km\,s$^{-1}$ with younger, more rapidly rotating stars having hotter coronae and faster winds. The mass loss rates can range from $10^{-14}-10^{-12}\,\mathrm{M}_\odot\mathrm{\,yr}^{-1}$. Coronal winds are composed of a mix of ions, electrons, and neutral atoms, primarily hydrogen and helium. 

\red{\cite{Waldron1985} and \cite{deJager1988} have derived empirical relations for mass loss rates across the HR diagram. These have been combined with the aforementioned rates for hot stars by \cite{Glebbeek2009}, and subsequently in the well-known Modules for Experimentation in Stellar Astrophysics (MESA) code ~\citep{Paxton2011,Paxton2013,Paxton2015}\footnote{In MESA this is referred to as the `Dutch' wind mass loss prescription.}.
It should be noted that this approach is not based on a physical model of the wind acceleration but rather on empirical relations.}

\subsection{Stars that have aged off the main sequence}\label{sec:evolved-type-stars}
Low to intermediate-mass \(M\lesssim 8 M_\odot\) stars evolve into asymptotic giant branch (AGB) stars, while massive stars \(M\gtrsim 10M_\odot\) evolve into red supergiant stars (RSG). 

The winds of evolved stars
are driven by the combination of stellar pulsations and radiation pressure on dust grains, which form in these stars' cool, extended atmospheres. Stellar pulsations produce shock waves that lift material away from the stars' surface. Dust grains, which form in the cool outer regions of the atmospheres, are pushed outwards by radiation pressure, dragging gaseous matter with them (see the reviews by~\citealt{Hofner2018,2021ARA&A..59..337D}). For evolved stars the wind velocities are low, ranging from 10--50\,km\,s$^{-1}$\red{,} but mass loss rates are high \red{(}$10^{-7}-10^{-4}\,\mathrm{M}_\odot\mathrm{\,yr}^{-1}$\red{)}. The winds of these evolved stars are rich in molecules and dust grains with primarily carbon- and oxygen-rich compounds.

The wind velocity profile for evolved stars is also often assumed to be that of a \(\beta\)-law as for massive stars (Equation~\ref{eq:beta-law}). The mass loss rate is different from massive stars, however.
For red supergiant stars,~\citet{1975MSRSL...8..369R} gave a semi-empirical law that describes the mass loss rate of a star as a function of its luminosity and radius:
\begin{equation}
\dot M = \eta \big(L\big/L_\odot\big)\big(R\big/R_\odot\big) \big(M\big/M_\odot\big)^{-1}
\end{equation}
where \(\eta=4\times10^{-13}\mathrm{\,M}_\odot\mathrm{\,yr}\); this has become known as the `Reimers relation'.
A different version \red{from}~\cite{2005ApJ...630L..73S} is
\begin{equation}
    \dot M = \eta \big(L\big/L_\odot\big)\big(R\big/R_\odot\big)  \big(M\big/M_\odot\big)^{-1}
    \left(\frac{T_\text{eff}}{4000\mathrm{\,K}}\right)^{3.5}
    \left(1+\frac{g_\odot}{4300g}\right)
\end{equation}
where \(\eta=8\times10^{-14}\mathrm{\,M}_\odot\mathrm{\,yr}\)

\end{appendices}


\bibliography{sn-bibliography}

\end{document}